\documentclass[]{aa}

\usepackage{graphicx}
\usepackage{natbib}
\usepackage{subfigure}

\bibpunct{(}{)}{;}{a}{}{,} 

\begin{document}

\title{The ESO-Spitzer Imaging extragalactic Survey (ESIS) I:\\
WFI B,V,R deep observations of ELAIS-S1 and comparison to Spitzer and GALEX 
data
\thanks{Based on observations collected at the European Southern
Observatory, Chile, ESO No. 168.A-0322(A).} \thanks{ESIS web page:
http://dipastro.pd.astro.it/esis/}}


\author{Stefano Berta\inst{1}
\and Stefano Rubele\inst{1} 
\and Alberto Franceschini\inst{1}
\and Enrico~V. Held\inst{2}
\and Luca Rizzi\inst{3}
\and Carol~J. Lonsdale\inst{4}
\and Thomas H. Jarrett\inst{4}		
\and Giulia Rodighiero\inst{1} 		
\and Seb~J. Oliver\inst{5} 
\and Joao~E. Dias\inst{6} 
\and Helen~J. Buttery\inst{6}
\and Fabrizio Fiore\inst{7} 
\and Fabio La Franca\inst{8}
\and Simonetta Puccetti\inst{7}	
\and Fan Fang\inst{4}
\and David Shupe\inst{4}
\and Jason Surace\inst{4}
\and Carlotta Gruppioni\inst{9}
}

\institute{Dipartimento di Astronomia, Universit\`a di Padova, 
Vicolo dell'Osservatorio 2, 35122 Padova, Italy
\and INAF -- Osservatorio Astronomico di Padova, Vicolo dell'Osservatorio 5,
35122 Padova, Italy
\and Institute for Astronomy, University of Hawaii, 2680 Woodlawn Drive, 
Honolulu, HI 96822, USA
\and Infrared Processing \& Analysis Center, California Institute of Technology
100-22, Pasadena, CA 91125, USA 
\and Astronomy Centre, CPES, University of Sussex, Falmer, Brighton BN19QJ, UK 
\and INAF -- Osservatorio Astronomico di Arcetri, Largo E. Fermi 5, 
50125 Firenze, Italy 
\and INAF -- Osservatorio Astronomico di Roma, Via Frascati 33, I-00044 
Monteporzio Catone, Italy 
\and Dipartimento di Fisica, Universit\`a\ degli Studi `Roma Tre', Via della Vasca 
Navale 84, I-00146 Roma, Italy 
\and INAF -- Osservatorio Astronomico di Bologna, via Ranzani 1, 
I-40127 Bologna, Italy 
}

\offprints{Stefano Berta, \email{berta@pd.astro.it}}

\date{Received ... / Accepted ...}

\titlerunning{ESIS}
\authorrunning{Berta S., et al. }

\abstract{
The \begin{em}ESO-Spitzer extragalactic Imaging Survey\end{em} (ESIS) is the optical follow up of 
the \begin{em}Spitzer Wide-Area InfraRed Extragalactic\end{em} (SWIRE) survey in the
ELAIS-S1 area.}{The multiwavelength study of galaxy emission is the key to
understand the interplay of the various components of galaxies and to trace
their role in cosmic evolution. ESIS provides optical identification and colors
of Spitzer IR galaxies and builds the bases for photometric redshift estimates.}
{This paper presents B, V, R Wide Field Imager observations of
the first 1.5 square degree of the ESIS survey.
Data reduction is 
described including astrometric calibration, illumination and color
corrections. Synthetic sources are simulated in scientific and super-sky-flat
images, with the purpose of estimating completeness and photometric accuracy for
the survey. 
Number counts and color distributions are compared to literature observational and
theoretical data, including non-evolutionary, PLE, evolutionary and
semi-analytic $\Lambda$CDM galaxy models,
as well as Milky Way stellar predictions.
The ELAIS-S1 area benefits
from extensive follow-up from X-ray to radio frequencies:
some potential uses of the multi-wavelength observations are illustrated.
}{Object coordinates are defined with an accuracy as good as $\sim
0.15$ $[$arcsec$]$ r.m.s. with respect to GSC 2.2; flux uncertainties are 
$\sim$2, 10, 20\% at mag. 20, 23, 24
respectively (Vega); we reach 95\% completeness at B,V$\sim$25 and R$\sim$24.5.
ESIS galaxy number counts are in good agreement with
previous works and are best reproduced by evolutionary and hierarchical
$\Lambda$CDM scenarios.  
Optical-Spitzer color-color plots promise to be very powerful tools to
disentangle different classes of sources (e.g. AGNs, starbursts, quiescent
galaxies). Ultraviolet GALEX data are matched to optical and Spitzer
samples, leading to a discussion of galaxy properties in the UV-to-24
$\mu$m color space. 
The spectral energy distribution of a few objects, from the X-rays to the
far-IR are presented as examples of the multi-wavelength study of galaxy
emission components in different spectral domains.}{}


\maketitle


\section{Introduction}

The assembly and evolution of galaxies, galaxy clusters
and cosmic large-scale structure (LSS) are currently major issues in both
theoretical and observational cosmology.
Tracing the history of cosmic star formation and the growth of the cosmic
stellar mass density will lead to an understanding of the
fundamental processes transforming the primordial diffuse plasma into the highly
structured present-day Universe. 

The multi-wavelength study of the emission of galaxies provides
tools to disentangle their different physical components.
Young stellar populations power the UV-optical light,
old stars emit predominantly in the near-IR, while dust --- heated by either starburst activity or
an active galactic nucleus (AGN) --- dominates the mid- and far-IR luminosity. X-ray and radio luminosities are
also produced by starbursts or AGNs.
Extending the analysis of galaxy properties to as wide a redshift range as
possible is compulsory, in order to understand the interplay of the various
components and to trace their role in cosmic galaxy evolution.

The two main instruments onboard Spitzer, observing in the
mid- to far-IR ($3-8\ \mu$m for IRAC and 24, 70, 160 $\mu$m for MIPS), were specifically
designed to probe the old stellar content and dust re-radiation 
from distant galaxies.

The cosmic Infra-Red Background \citep[CIRB,][]{puget1996,hauser1998},
discovered by the COBE satellite in the mid 90's, 
is the most energetic diffuse radiation after the CMB. 
\citet{elbaz2002} and \citet{franceschini2003} have shown that at least 50\% of the CIRB emission
is powered by luminous and ultra-luminous massive star-forming galaxies, at
redshift $z=0.5-1.5$, strongly evolving in cosmic time.
The Universe seems to
have experienced a phase of enhanced activity of star-formation
and gravitational accretion in the past, mostly visible in the infrared
\citep[e.g.][]{franceschini2001}. 

The Spitzer {\em Multiband Imaging Photometer} \citep[MIPS,][]{rieke2004} provides the observations
needed to push the study of 
mid- and far-IR sources to fainter luminosities and larger distances. Polycyclic Aromatic
Hydrocarbon (PAH) emission, typical of starburst 7-13 $\mu$m restframe spectra, is
sampled by MIPS up to redshift $\sim 3$, corresponding to an epoch when the Universe
had only $\sim$15\% of today's age.

A central issue in modern cosmology is 
{\em when} galaxies assembled their baryonic mass. 
The
hierarchical scenario \citep[e.g.][]{kauffmann1998} predicts that the most massive systems (e.g.
$M_{\textrm{stars}} > 10^{11}$ M$_\odot$) formed relatively late through a slow
process of merging of smaller galaxies. 
In monolithic-collapse models \citep{eggen1962}, the bulk of 
stars formed in early-type galaxies at very high redshift, while subsequent
merging and star formation are limited. Modern $\Lambda$CDM models and hydrodynamical 
simulations \citep[e.g.][]{nagamine2001} implement a mixture of the two.

Several studies have attempted to investigate the formation and
evolution of massive systems and test model predictions 
\citep[see, e.g., the recent works by][]{drory2005,treu2004,fontana2004,cimatti2003},  
but a clear and coherent picture has not yet emerged.
Very little is known about galaxy stellar mass assembly at redshift $z > 1.5$
\citep[e.g.][]{cimatti2003,daddi2004}. 

The {\em Infrared Array Camera} \citep[IRAC,][]{fazio2004} onboard Spitzer observes 
in the $3.6-8.0$ $\mu$m wavelength range.  
The instrument was specifically designed for detecting galaxies' restframe
near-IR emission, up to redshift $z\ge3$, hence directly 
probing their stellar mass assembly.

\subsection{The SWIRE \& ESIS Surveys}

A significant fraction of the first year of Spitzer in-flight operations has been devoted to six different
Legacy science Programs, representing projects of general and lasting importance
to the broad astronomical community.
Among these, SWIRE \citep{lonsdale2003,lonsdale2004} and 
GOODS \citep{dickinson2003} are dedicated to cosmology.

The {\em Spitzer Wide-area Infra-Red Extragalactic} survey
(SWIRE\footnote{SWIRE web page: http://swire.ipac.caltech.edu/}) 
is the largest 
Spitzer Legacy Program. It consists of a wide-area, imaging campaign designed to trace the
evolution of dusty, star-forming galaxies, evolved stellar populations, and AGN
as a function of environment, from redshifts $z\sim3$, down to the current
epoch. SWIRE includes 6 high-latitude fields, totaling 49 $[$deg$^2]$ in all
the seven Spitzer bands.

The large sky area covered by the SWIRE survey implies large galaxy samples 
of all kinds, secures high statistical significance to clustering and galaxy 
evolution studies, and allows the possibility of rare object searches. 

The {\em ESO-Spitzer wide-area Imaging Survey} (ESIS) is an ESO Large Programme (P.I.
Alberto Franceschini), securing optical
ground-based imaging follow-up to the SWIRE Spitzer survey 
in the ELAIS-S1 field. 

The ELAIS-S1 region, together with Lockman Hole, is
the highest priority region for SWIRE, thanks to the extremely low 100 $\mu$m
cirrus emission \citep{schlegel1998}.
In fact, it includes the absolute minimum of the Galactic 100 $\mu$m
emission in the Southern sky. 

A broad wavelength coverage is required by the nature of the
sources we are targeting, which are typically reddened, and/or with old evolved stellar
populations, and at high redshifts. These properties imply dramatic
difficulties when trying to acquire spectroscopic follow-up in the optical. 
Photometric investigation of the properties of faint sources provides
a very powerful alternative, and it is clearly the only viable approach when
dealing with large datasets over wide areas.

\begin{figure}[!t]
\centering
\includegraphics[width=0.45\textwidth]{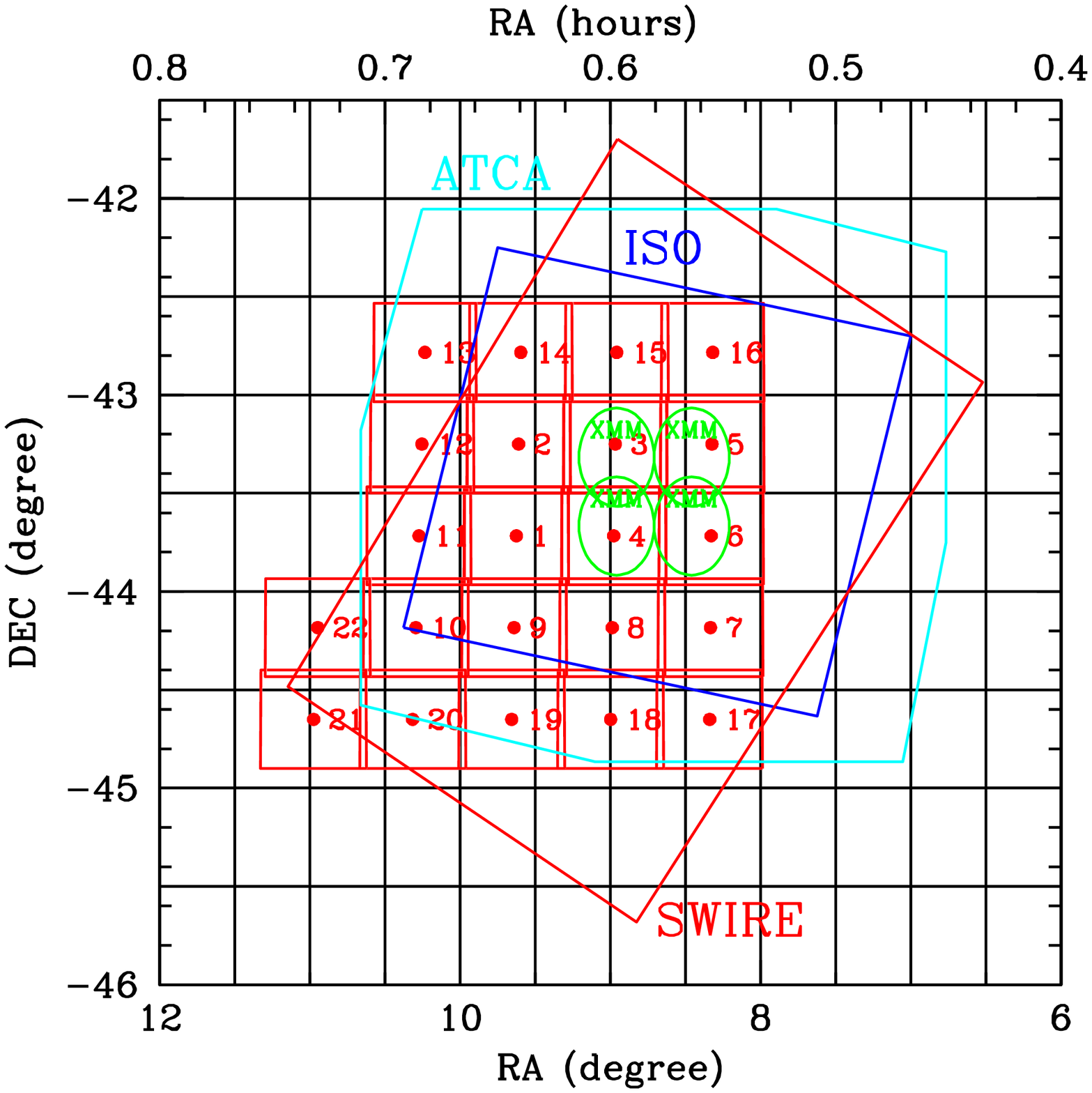}
\caption[WFI ESIS coverage]{ESIS WFI (BVR) pointings
scheduled in the ELAIS S1 region: a total of 22 fields (red numbered small squares) of 
$30'\times30'$ are being observed as part of the ESIS survey.
The region observed by ISO \citep{oliver2000}, the Spitzer/SWIRE field 
\citep{lonsdale2003} and the ATCA radio survey 
by \citet{gruppioni1999} are marked. The four circles represent XMM 
observations.
Regions 3-6 have also been observed with SOFI/NTT with J and Ks filters, during
Summer 2002. See Section \ref{sec:multiwave} for a review of the available 
data in ELAIS-S1.}
\label{fig:wfi_strategy}
\end{figure}

ESIS covers $\sim 5$ $[$deg$^2]$ in 5 optical
bands and is based on imaging with the {\em Wide Field Imager} \citep[WFI, at the focal plane of the 2.2m La
Silla ESO-MPI telescope,][]{baade1999} and the {\em VIsible Multi Object
Spectrograph} \citep[VIMOS, on VLT,][]{lefevre2002,dodorico2003} to $\sim25-26$
mag in BVRIz. 
The total amount of scheduled observing time is 27
nights with the WFI and 8 nights with VIMOS. 

Prime motivations for ESIS are to: 
\begin{itemize}
\item obtain optical identification for the roughly 300000 IR sources detected
by SWIRE in the 5 $[$deg$^2]$ of the ESIS area; the present statistics on
Spitzer sources indicate that $\sim80$\% of the SWIRE IRAC
sources could be detected to $B=26$ and $V=25.5$; 
\item provide colors and rough morphologies for source classification;
\item build the basis for photometric redshifts for all IR sources, and optimize the
spectroscopic IR and optical follow-up; 
\item provide UV-blue restframe luminosities of IR galaxies;
\item produce independent optical samples, selected on the basis of their
restframe UV-blue properties, for comparison with the IR-selected ones.
\end{itemize}

This paper describes WFI observations, data reduction and analysis in the 
central ELAIS-S1 1.5 $[$deg$^2]$. VIMOS observations will be characterized in a 
forthcoming paper and subsequent releases of WFI data will be presented on the
ESIS official web page.
The data described here
are being included in the third SWIRE Spitzer Legacy data release (Fall 2005).

Section 2 deals with 
optical Wide Field Imager service mode observations. Data reduction is
described in Section 3, where astrometric accuracy, catalog
extraction and photometric calibration are also discussed.
In Section 4, we present a set of simulations carried out with the purpose of
estimating the effective depth of the survey and flux uncertainties. 
Section 5 deals with source number counts, disentangling point-like and
extended objects and comparing ESIS to literature data and predictions of galaxy
evolutionary models. In Section 6, the ESIS catalog is matched to
multiwavelength data in the ELAIS-S1 area, spanning from the X-rays to the
far-infrared, illustrating the potential of panchromatic studies of galaxies.
Finally, in Section 7, we summarize current findings.

\begin{figure}[!t]
\centering
\rotatebox{-90}{\includegraphics[height=0.45\textwidth]{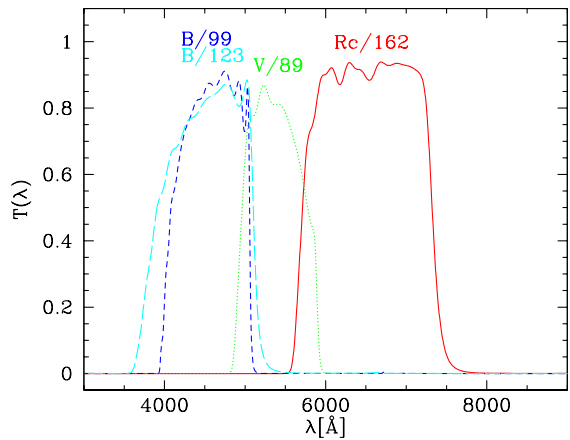}}
\caption{Transmission curves for ESIS WFI observations.}
\label{fig:filters}
\end{figure}

\begin{table}[!t]
\centering
\begin{tabular}{ccc}
\hline
\hline
Exp. \# & $\Delta\alpha$ & $\Delta \delta$ \\
\hline
1 & 0 & 0 \\
2 & +32 & +24 \\
3 & --32 & +48 \\
4 & --48 & --24 \\
5 & +48 & --48 \\
\hline
\end{tabular}
\caption{Dithering pattern in $[$arcsec$]$ of WFI BVR observations. Each pointing consists of 
30 dithered exposures (6 sets of 5).
The shifts reported here refer to the first exposure of each set.}
\label{tab:dither}
\end{table}

\begin{table*}[!t]
\centering
\tiny
\begin{tabular}{c c c| c c c | c c c | c c c}
\hline
\hline
\multicolumn{3}{c|}{Field} & \multicolumn{3}{c|}{B band}  & \multicolumn{3}{c|}{V band} & \multicolumn{3}{c}{R band} \\
\hline
Exposure & $\alpha$ & $\delta$ & Date & Airmass & Seeing & Date & Airmass & Seeing & Date & Airmass &
Seeing\\
set & $[$hh:mm:ss$]$ & $[^\circ:':'']$ & dd/mm/yy & (avg) & $['']$ & dd/mm/yy & (avg) & $['']$ & dd/mm/yy &
(avg) & $['']$ \\
\hline
\hline
1\_1 & 00:38:30.0 & -43:43:00 & 10/10/01 & 1.338 & 0.81 & 10/10/01 & 1.502 & 0.83 & 10/10/01 & 1.222 & 0.71 \\
1\_2 &		  &	      & 11/10/01 & 1.313 & 0.86 & 24/10/01 & 1.034 & 1.07 & 24/10/01 & 1.210 & 0.92 \\
1\_3 &		  &	      & 11/10/01 & 1.153 & 0.70 & 17/11/01 & 1.048 & 1.05 & 24/10/01 & 1.131 & 0.98 \\
1\_4 &		  &	      & 12/10/01 & 1.281 & 1.09 & 17/11/01 & 1.034 & 0.83 & 24/10/01 & 1.080 & 0.88 \\
1\_5 &		  &	      & 12/10/01 & 1.182 & 1.10 & 17/11/01 & 1.035 & 0.84 & 22/10/01 & 1.049 & 0.82 \\
1\_6 &		  &	      & 13/10/01 & 1.283 & 1.08 & 14/12/01 & 1.133 & 0.89 & 24/10/01 & 1.083 & 1.06\\
\hline
2\_1 & 00:38:26.4 & -43:15:00 & 06/11/01 & 1.063 & 0.98 & 06/11/01 & 1.048 & 1.03 & 06/11/01 & 1.143 & 0.97 \\
2\_2 &		  &	      & 24/10/01 & 1.174 & 1.02 & 24/10/01 & 1.084 & 1.10 & 24/10/01 & 1.048 & 1.08 \\
2\_3 &		  &	      & 09/07/02 & 1.040 & 0.82 & 08/07/02 & 1.038 & 0.91 & 10/08/02 & 1.042 & 0.80 \\
2\_4 &		  &	      & 09/07/02 & 1.030 & 0.88 & 10/07/02 & 1.032 & 1.00 & 14/08/02 & 1.126 & 0.87 \\
2\_5 &		  &	      & 11/07/02 & 1.148 & 1.05 & 11/07/02 & 1.053 & 0.95 & 14/08/02 & 1.043 & 0.72 \\
2\_6 &		  &	      & 11/07/02 & 1.229 & 1.09 & 11/07/02 & 1.090 & 0.96 & 14/08/02 & 1.149 & 0.71 \\
\hline
3\_1 & 00:35:52.0 & -43:15:00 & 11/10/01 & 1.479 & 0.91 & 11/10/01 & 1.503 & 1.01 & 22/10/01 & 1.170 & 0.72 \\
3\_2 &		  &	      & 06/11/01 & 1.032 & 1.02 & 06/11/01 & 1.086 & 1.02 & 06/11/01 & 1.236 & 1.07 \\
3\_3 &		  &	      & 13/10/01 & 1.147 & 1.04 & 30/09/02 & 1.042 & 1.06 & 30/09/02 & 1.061 & 0.78 \\
3\_4 &		  &	      & 20/10/03 & 1.108 & 0.91 & 30/09/02 & 1.031 & 1.00 & 18/11/01 & 1.031 & 1.05 \\
3\_5 &		  &	      & 10/10/01 & 1.277 & 0.80 & 30/09/02 & 1.036 & 0.83 & 30/09/02 & 1.104 & 0.93 \\
3\_6 &		  &	      & 09/10/01 & 1.271 & 0.97 & 30/10/03 & 1.050 & 0.90 & 20/10/03 & 1.035 & 0.71 \\
\hline
4\_1 & 00:35:54.4 & -43:43:00 & 28/11/03 & 1.038 & 0.99 & 28/10/02 & 1.064 & 1.18 & 28/10/02 & 1.109 & 1.04 \\
4\_2 &		  &	      & 28/11/03 & 1.182 & 1.00 & 11/07/02 & 1.035 & 1.13 & 20/10/03 & 1.149 & 0.72 \\
4\_3 &		  &	      & 30/10/03 & 1.102 & 0.73 & 11/07/02 & 1.034 & 1.03 & 30/10/03 & 1.138 & 0.75 \\
4\_4 &		  &	      & 01/11/03 & 1.036 & 0.98 & 20/10/03 & 1.089 & 0.77 & 30/10/03 & 1.219 & 0.92 \\
4\_5 &		  &	      & 30/10/03 & 1.129 & 0.94 & 30/10/03 & 1.066 & 0.75 & 01/11/03 & 1.075 & 0.89 \\
4\_6 &		  &	      & 20/10/03 & 1.040 & 0.94 & 12/07/02 & 1.122 & 1.10 & 30/10/03 & 1.033 & 0.98 \\
\hline
5\_1 & 00:33:17.6 & -43:15:00 & 27/10/02 & 1.071 & 1.02 & 27/10/02 & 1.120 & 0.97 & 27/10/02 & 1.193 & 1.02 \\
5\_2 &		  &	      & 30/11/03 & 1.140 & 0.77 & 30/11/03 & 1.047 & 0.83 & 30/11/03 & 1.082 & 0.90 \\
5\_3 &		  &	      & 28/11/03 & 1.388 & 1.04 & 28/11/03 & 1.255 & 0.95 & 30/11/03 & 1.226 & 0.76 \\
5\_4 &		  &	      & 05/01/05 & 1.535 & 0.91 & 30/11/03 & 1.418 & 0.85 & 10/01/05 & 1.521 & 0.94 \\
5\_5 &		  &	      & 	 &       & & 	      &       & &	   &	   & \\
5\_6 &		  &	      & 20/10/03 & 1.141 & 0.70 & 20/10/03 & 1.233 & 0.80 & 20/10/03 & 1.383 & 0.72 \\
\hline
6\_1 & 00:33:18.8 & -43:43:00 & 	 &       & & 30/10/02 & 1.065 & 1.25 &	   &	   & \\
6\_2 &		  &	      & 30/10/02 & 1.040 & 1.22 & 	      &       & & 30/10/02 & 1.110 & 1.30 \\
6\_3 &		  &	      & 	 &       & & 	      &       & &	   &	   & \\
6\_4 &		  &	      & 	 &       & & 	      &       & &	   &	   & \\
6\_5 &		  &	      & 	 &       & & 	      &       & &	   &	   & \\
6\_6 &		  &	      & 21/10/03 & 1.086 & 0.70 & 21/10/03 & 1.222 & 0.70 & 21/10/03 & 1.141 & 0.78 \\
\hline
\end{tabular}
\footnotesize
\caption{Summary of WFI, BVR observations. Each set of exposures consists of
five 300 $[$s$]$ frames. Observing dates and average airmasses are reported for
each set. Seeing values have been measured directly on science frames.}
\normalsize
\label{tab:observations}
\end{table*}

\section{Observations}\label{sec:observations}

The ESIS WFI survey consists of BVR imaging observations of ELAIS-S1, carried 
out in {\em Service} mode between October 2001 and 
January 2005. Further observations are scheduled and will be continued
until completion of the entire 5 $[$deg$^2]$ planned area. 

The ESIS field has been divided into 22 different sub-areas,
each corresponding to one WFI $30'\times30'$ pointing. 
The different pointings
overlap by about 1.5$'$ on each side to obtain a full coverage and
to allow trimming individual images to avoid edge effects.

Figure \ref{fig:wfi_strategy} shows the WFI coverage of the ELAIS-S1 ESIS sky 
area, as well as the SWIRE, ISO, XMM, ATCA fields. The SWIRE and ESIS
fields are shifted relative to one another because of resizing of the original Spitzer
area.
Section \ref{sec:multiwave} will describe the available multi-wavelength 
observations in ELAIS-S1.

The WFI array is made of 8 CCDs, separated by small gaps. In order to fill 
these gaps and correct bad pixels and cosmic ray events, 
each $30'\times30'$ field is covered by 6 sets ({\em Observing Blocks}, OBs) of 5 
dithered exposures. At least one of the six OBs per pointing per band was
observed during a photometric night, as required by the scheduling.
The standard WFI dithering pattern (see Tab. \ref{tab:dither}) was used. 
The scale of the instrument is 0.238 $[$arcsec/pixel$]$.

Each single exposure consists of 300 $[$s$]$, for a total exposure time
of 2.5 hours per pointing per band (B, V, R).

The filters used are WFI B/99, V/89 and Rc/162. Observing Blocks defined after
September 2002 (i.e. starting from pointing no. 5) have B/123 instead of B/99,
because the standard WFI filters and calibration plan changed; in this paper we
refer to B,V,R, regardless of which are involved. There are
no significant differences in the final catalog, which is calibrated to the
Johnson-Cousins system (see Sect. \ref{sec:phot}). Figure \ref{fig:filters} shows the
transmission functions for the 4 filters.

Table \ref{tab:observations} summarizes the observing logs 
for the central 6 fields (i.e. 1.5 $[$deg$^2]$). Field 6 observations have not
been completed yet, but this field is included in this paper because it covers part of
the area observed in the X-rays and near-IR. The seeing and airmass values are
the averages on the 5 images constituting each OB. 
The majority of nights satisfied the requested observing constraints: 
seeing$<$1.2\arcsec, moon distance $>$120$^\circ$, moon phase $<0.4$. Aborted, or 
out-of-requirements OBs are not listed in Table \ref{tab:observations}, apart 
from those for pointing no. 6.

\section{Data Reduction}

The reduction of ESIS data was performed within the IRAF\footnote{The
package IRAF is distributed by the National Optical Astronomy Observatory which
is operated by the Association of Universities for Research in Astronomy, Inc.,
under cooperative agreement with the National Science Foundation.} environment,
using the package WFPRED developed by two of us (LR, EVH) at the Padova
Astronomical Observatory. This package consists of a set of procedures
that access and upgrade the standard MSCRED \citep{valdes1998} tasks.

Correction of bias and flat-fielding followed the standard procedure.
Sky flat-field frames were acquired during each night and applied to images
obtained on the same date.

\subsection{Super-Sky-Flat Field}
Even if sky flat-field frames were used, significant 
illumination gradients exist on the pre-reduced frames.
Hence a further correction was needed, both to obtain a uniform background and
to build frames with the same photometric zeropoint across the whole field of
view. 
The main causes for these gradients are:
\begin{enumerate}
\item sky flat fields point at different positions on the
sky than the science target, and therefore have different relative positions
to the moon and very bright stars;
\item bright off-axis sources produce spurious multiple reflections in the
focal reducer of WFI;
\item the moon, when present, produces smooth gradients in the background of
science frames.
\end{enumerate}
To address these issues, Super-sky-flat field frames (hereafter {\em ss-flat}) 
were produced, 
by combining all science frames obtained during a given night. The adopted 
procedure is sketched below.
\begin{figure}[!t]
\centering
\includegraphics[width=0.45\textwidth]{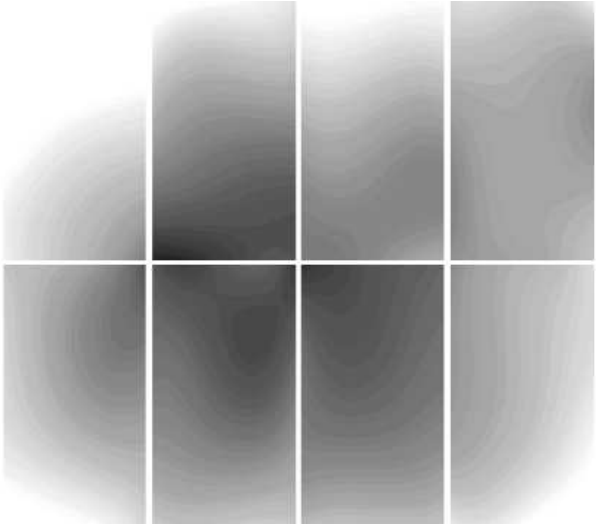}
\caption{Example of R-band super-sky-flat frame (negative rendered).}
\label{fig:ss_flat}
\end{figure}
\begin{itemize}
\item In order to create the {\em ss-flat}, astronomical objects must be removed from
the science images. 
Detection of all sources and build up of {\em object masks} were performed
with the {\tt objmasks} task in the MSCRED IRAF package.
\item Very bright sources (e.g. luminous stars and big galaxies) produce very
extended halos on the images. All science frames were visually inspected and
all halos were manually masked.
\item All science frames were then combined together, using object-masks to
cancel-out the light coming from astronomical sources.
\item The result was fitted with a 2-dimensional 4th order Legendre function, 
obtaining a smoothed, but accurate representation of the {\em ss-flat}. Alternatively,
a simple boxcar smoothing was also tested, but it was found that Legendre fitting
led to better results, because it avoids ``holes'' produced by the large masks
on extended halos.
\end{itemize}
An example of R band {\em ss-flat} is shown in Figure \ref{fig:ss_flat}.

Since the targeted fields lie all roughly at the same equatorial coordinates,
the same {\em ss-flat} was applied to all the images taken during a given night,
unless significant changes in sky conditions (e.g. presence of
atmospheric cirrus) occurred.

\begin{figure}[!t]
\centering
\includegraphics[width=0.405\textwidth]{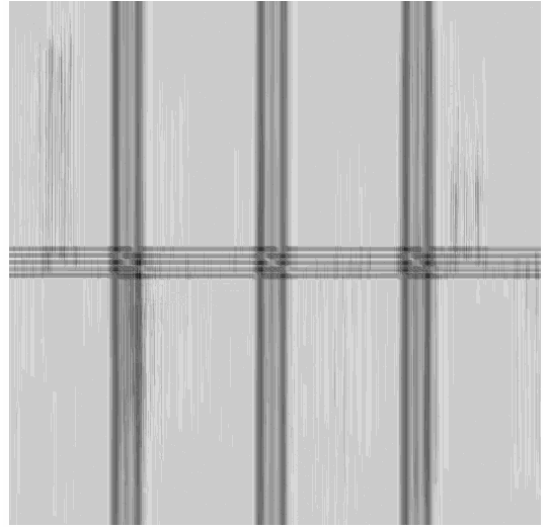}
\caption{Weight mask for R-band ESIS WFI pointing no. 2, used during the detection 
of sources and the estimation of photometric uncertainties.}
\label{fig:weights}
\end{figure}

\subsection{Astrometric calibration and co-addition}\label{sec:astrom}
Reduction of wide field imaging data requires a detailed astrometric 
calibration, in order to take into account the
intrinsic distortions of the instrumentation, and minimize the effect of 
projecting a large sky area (intrinsically spherical) onto a planar CCD.

In the case of WFI ESIS data, the $30'\times30'$ field of view requires
particular care when mapping pixels into celestial coordinates.
In view of multi-wavelength cross-correlation and future spectroscopic
follow up in the ELAIS-S1 area, as good an astrometric calibration as 
possible is necessary. 

An accurate solution is provided by the TNX transformation, that combines a
linear projection of the sky sphere onto a tangent plane 
\citep[the standard {\em gnomonic} algorithm, TAN,][]{calabretta2002} and a
polynomial function for distortions. A {\em simplified} description of the
adopted mapping is:
\begin{eqnarray}
\left( \begin{array}{c} \xi'\\ \eta' \end{array} \right) &=&
\left( \begin{array}{c} \xi\\ \eta \end{array} \right) +
\left( \begin{array}{c} lngcor[\xi,\eta]\\ latcor[\xi,\eta] \end{array} \right)\\
\left( \begin{array}{c} \xi\\ \eta \end{array} \right) &=&
\left( \begin{array}{cc} a_{11} & a_{12}\\ a_{21} & a_{22}\end{array} \right)
\left( \begin{array}{c} x-x_c \\ y-y_c \end{array} \right)
\end{eqnarray}
where $(\xi',\eta')$ are the celestial coordinates on the sky, mapped by the TNX
algorithm,  $(\xi,\eta)$ represent the first-order solution of the TAN
transformation (given by the $a$ matrix), $(x,y)$ and $(x_c,y_c)$ are the
cartesian coordinates of the generic pixel and of the image center,
$lngcor$ and $latcor$ are the polynomial equations mapping distortions
along longitude and latitude.

\begin{figure*}[!t]
\centering
\includegraphics[width=0.85\textwidth]{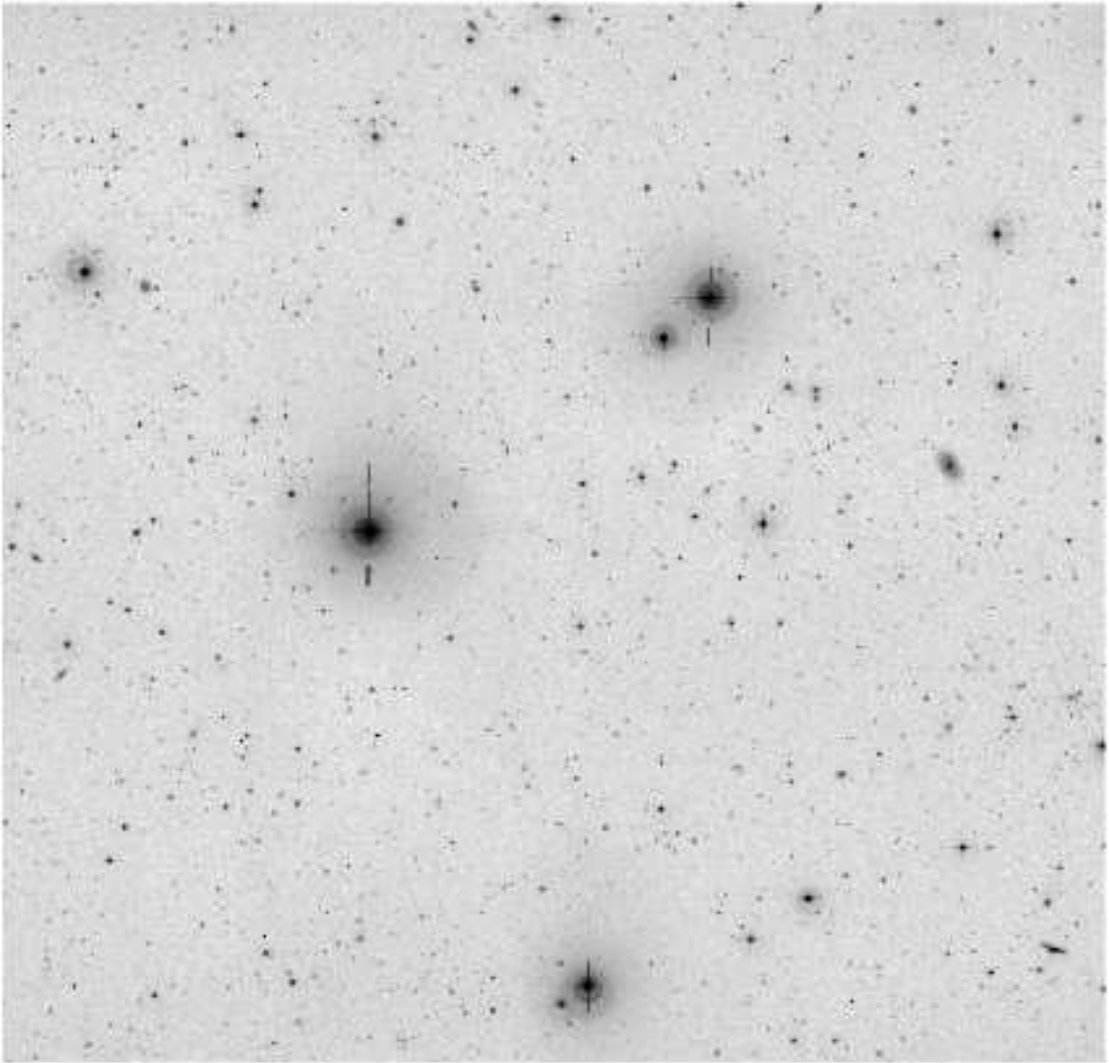} 
\caption{Final, co-added, R-band frame of ESIS WFI Field number 3.}
\label{fig:field3}
\end{figure*}

The solution of the above equation array is built on dedicated observations of
\citet{stone1999} astrometric fields D and E, carried out as part of the ESIS program. 
Since observations are spread along a wide period of time, several different
astrometric solutions have been necessarily built, because extra-ordinary
maintenance of the WFI camera affected the {\em pixels}$\rightarrow${\em
celestial} coordinates map.

The best results were obtained using 4th order distortion polynomials. Every
individual image was astrometrically calibrated before co-addition. 
After the solution was applied, frames were finely re-centered
by means of rigid shifts to the GSC 2.2 (STScI \& OaTO, \citeyear{gsc2}) 
catalog.

During co-addition, all frames were re-projected, transforming their astrometric
map to a common one (that for the B band), regardless of the photometric
band. 
This choice allows the B, V, and R images to be easily compared and 3-band catalogs
to be straightforwardly produced.

\begin{figure*}[!ht]
\centering
\includegraphics[width=0.88\textwidth]{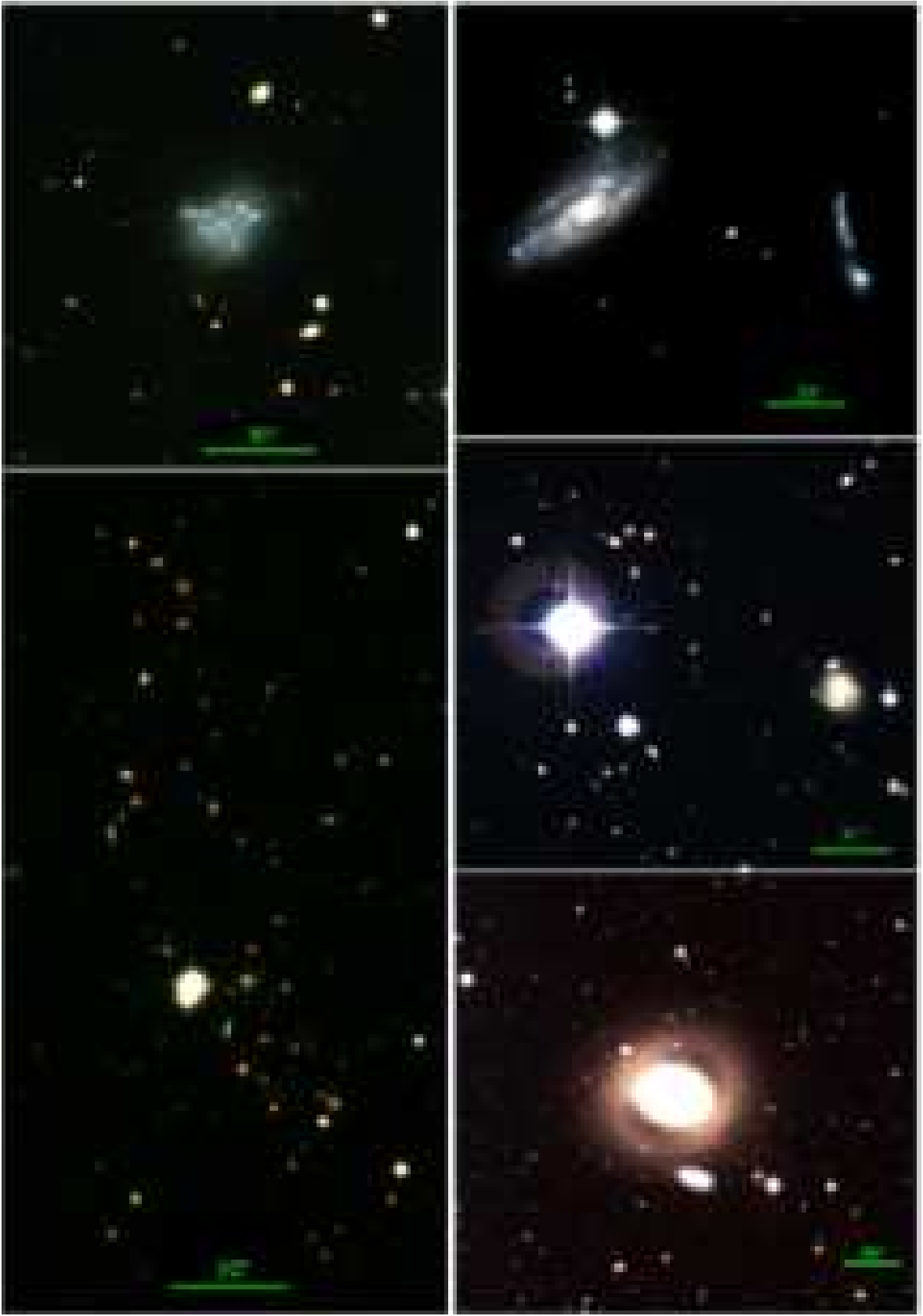}
\caption{Color, BVR, enlarged view of some sources in the ESIS fields.
North is up, East is left; a 30\arcsec scale is indicated.}
\label{fig:zooms_col}
\end{figure*}

Each one of the 22 ESIS WFI pointings consists of at least $\sim$30 science frames
per band (B,V,R, see Table \ref{tab:final_frames}). 
While co-adding all these frames together, we have excluded all those images
with seeing larger than $\sim 1.2$ $[$arcsec$]$, with the exception of pointing
no. 6, for which few images are available. The FWHM of point-like sources
(see next Sections), as measured by SExtractor \citep{bertin1996} on the final
images, is reported in the fourth column of Tab. \ref{tab:final_frames}. 

Figure \ref{fig:weights} 
shows the {\em weight map} obtained during co-addition of Field no. 2, R
band. 
The darkest regions have about half the exposure coverage of the lightest
ones, due
to gaps between the CCDs and dithering. 
The weight value defined as the ratio
$N_{imgs}/N_{tot}$ between the effective number of images covering the given 
pixel and the total number of frames belonging to the given ESIS pointing.
The weight maps were used by
SExtractor in object detection and noise calculation.
Figure \ref{fig:field3} shows the R-band image of WFI
Field number 3; Figure \ref{fig:zooms_col} includes 
some zooms on colorful sources. 

The coordinates of the objects detected on the final co-added R-band image 
of one WFI field are
compared to GSC 2.2 sources in Fig. \ref{fig:astrom} (left panels).
The {\em r.m.s}
of the distribution of coordinate differences turns out to be $\sim0.12$
$[$arcsec$]$ in
$\alpha\cdot\cos\delta$ and $\sim0.10$ $[$arcsec$]$ in $\delta$, 
similar to what found by \citet{arnouts2001} and \citet{momany2001}. 
The top left panel
of Fig. \ref{fig:astrom} includes also comparison of data to Gaussian 
distributions with the same variance $\sigma$.
The central plot shows that 70, 80 and 90\% of the
sources are included within $\Delta$ of $\sim 0.13,\ 0.16$ and $0.20$ $[$arcsec$]$
respectively\footnote{defined as $\Delta=\sqrt{[\Delta(\alpha\
\cos\delta)]^2+[\Delta(\delta)]^2}$}. Similar results are obtained for the B and V 
bands and the other pointings. In the bottom graph, coordinate differences are plotted against
$(\alpha,\delta)$: no systematic trends are detected.

Relative astrometry (i.e. the difference between
coordinates in the 3 bands) is very accurate, 
because the three monochromatic frames were registered to the same 
astrometric map, as mentioned above. The right panels of Fig.
\ref{fig:astrom} show  
the behavior of B {\em vs.} R coordinates for one WFI pointing; {\em r.m.s} are
below 0.05 $[$arcsec$]$.
Similar results are obtained also for V band and the other fields.

The astrometric accuracy has been checked also in the regions of overlap
between ESIS fields. 
The {\em bottom right} panel of Fig. \ref{fig:astrom} reports the result for 
GSC 2.2 sources: coordinates measured on each pointing are compared to those
estimated on contiguous fields. The plot includes all the central six ESIS 
fields.

\subsection{Catalog extraction}\label{sec:extraction}

Source extraction and magnitude measurement were performed using the
SExtractor software \citep{bertin1996}. 

\begin{figure*}[!t]
\centering
\includegraphics[width=0.92\textwidth]{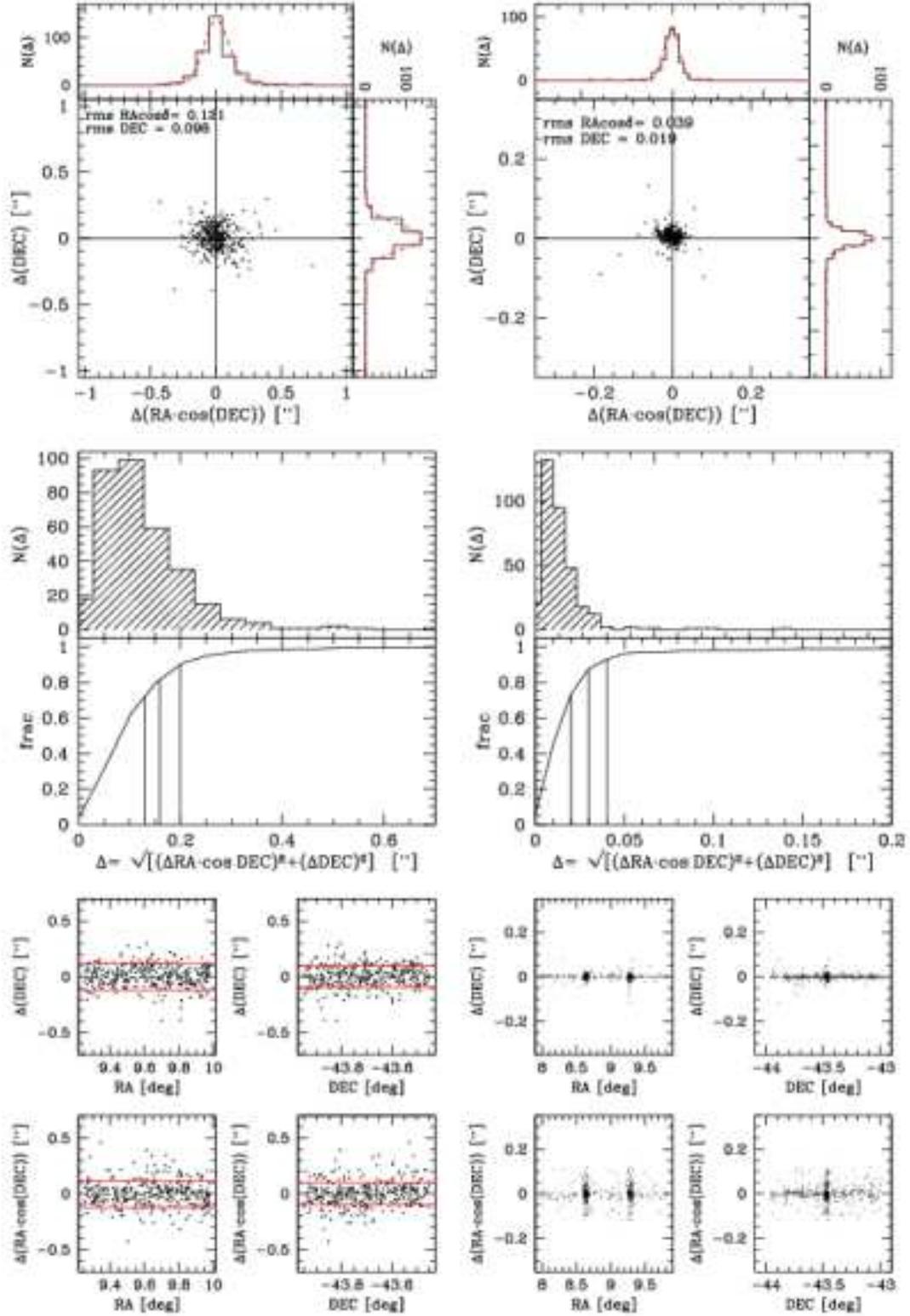}
\caption{Astrometric accuracy of ESIS WFI data.
{\em Left panels}: absolute R-band astrometric accuracy. The equatorial coordinates of
ESIS sources are compared to those in the GSC 2.2 catalog. The {\em central}
plot shows the distribution of $\sqrt{[\Delta(\alpha\cdot
\cos\delta)]^2+[\Delta(\delta)]^2}$; the
vertical lines include 70, 80 and 90\% of the sources respectively. In the {\em bottom
left} panel, the values of $\Delta(\alpha\cdot\cos\delta)$ and $\Delta(\delta)$
are plotted against $(\alpha,\delta)$.
{\em Right panels:} relative astrometric accuracy. The coordinates
measured on the B band image are compared to those in the R band. 
In the {\em bottom right} plot, the difference in coordinates among contiguous
fields is shown, for common GSC 2.2 sources.}
\label{fig:astrom}
\end{figure*}

Thanks to the astrometric transformations described in Section \ref{sec:astrom}, 
the B, V and R images of each WFI pointing are perfectly aligned (see Fig.
\ref{fig:astrom}) with each other. Therefore it is very easy to
cross-correlate single band source lists and generate a 3-band catalog.

We ran SExtractor on a B+V+R image obtained by simply summing the three
monochromatic frames belonging to each WFI field. A Gaussian-filtered 
image was used for detection, with a kernel matching the seeing.
We use this first 3$\sigma$ detection on the BVR image to build a list of
objects (ASSOC\_LIST) to be extracted and measured on the individual
monochromatic frames. In this way, spurious detections on individual images are
also minimized.

Photometry was then performed at 3$\sigma$ above background {\em r.m.s.} with
SExtractor also.
As a consequence of the dithering observing technique, regions along image
borders and CCD gaps are covered by a small number of frames; weight maps (see
Sect. \ref{sec:astrom}) are used by SExtractor while detecting sources and
computing magnitude uncertainties.
Nonetheless, it is worth noting that magnitude uncertainties ($<$0.1 mag down to
magnitude $\sim$25, for all the three bands) computed by
SExtractor likely underestimate the real errors in photometric
measurement, because they include only photon noise statistics 
(see tests in Section \ref{sec:phot_err}).

Table \ref{tab:final_frames} reports the number of sources detected and
measured by SExtractor on ESIS individual WFI frames. The total number of sources
in pointings 1 to 6, is 132712, smaller than the sum
on individual images because of overlap; 
the total number of objects on Fields 1-5 --- full depth --- is 118197.

\begin{table}[!t]
\centering
\footnotesize
\begin{tabular}{c c c c}
\hline
\hline
Field no. & Number & 3$\sigma$ & FWHM\\
\& filter & of frames & sources & [arcsec]\\
\hline
\hline
1 B & 30 & 22999 & 1.008 \\
1 V & 28 & 18381 & 0.936 \\
1 R & 32 & 24006 & 0.960 \\
\hline
2 B & 25 & 18623 & 0.984 \\
2 V & 26 & 11780 & 1.176 \\
2 R & 30 & 19501 & 0.960 \\
\hline
3 B & 25 & 20385 & 0.984 \\
3 V & 30 & 18108 & 0.960 \\
3 R & 32 & 26118 & 0.912 \\
\hline
4 B & 30 & 18186 & 0.984 \\
4 V & 30 & 16016 & 1.056 \\
4 R & 30 & 23754 & 0.960 \\
\hline
5 B & 25 & 19835 & 0.936 \\
5 V & 25 & 15991 & 0.912 \\
5 R & 25 & 22564 & 0.876 \\
\hline
6 B & 10 & 14916 & 0.852 \\
6 V & 10 & 10441 & 0.912 \\
6 R & 10 & 14720 & 0.864 \\
\hline
\end{tabular}
\normalsize
\caption{Summary of source extraction. For each of the 6 ESIS WFI pointings
analysed here, the total number of co-added frames, number of 3$\sigma$ extracted sources and
FWHM of point-like objects are reported.}
\label{tab:final_frames}
\end{table}

\subsection{Photometric calibration}\label{sec:phot}
The fundamental requirement to be fulfilled by the ESIS {\em service mode} WFI
program consisted of observing at least one OB (i.e. one set of 5 exposures,
hereafter ``{\em reference} OB'') per
pointing, per band, during one photometric night.
During the same photometric night, imaging of \citet{landolt1992} Ru149 or TPHE
spectrophotometric standard fields --- at similar airmasses --- was performed.
Eight different exposures were taken each time, in order to include the main
standard stars on each individual CCD of the WFI array.

When combining $\sim$30 different images, belonging to various nights (see
Tables \ref{tab:observations} and \ref{tab:final_frames}), frames
obtained with different sky conditions are mixed, hence the photometric 
information gets
lost. Having a photometric reference night is necessary, to recover a correct
calibration of magnitudes. The five images belonging to the {\em reference} OB were
combined together and were used to determine the zeropoint shift 
$\Delta$mag$=$mag$_{coadd}-$mag$_{ref}$, 
caused by the co-addition of $\sim$30 frames into the final frame. 

The preliminary catalog (Sect. \ref{sec:extraction}) was matched to the
photometric conditions and normalized to unit airmass and exposure time, 
using the equation: 
\begin{eqnarray}
\textrm{mag}_{phot}&=&\textrm{mag}_{coadd}+2.5\log_{10}(t_{exp})\\
\nonumber &&-K_\lambda \cdot A.M.-\Delta \textrm{mag}
\end{eqnarray}
where $A.M.$ is the average airmass of the reference OB, and 
$K_\lambda$ are the atmospheric extinction coefficients
provided by ESO ($K_{b99}=0.23$, $K_{b123}=0.22$, $K_v89=0.11$, 
$K_{r162}=0.07$).

After pre-reduction and {\em ss-flat} correction of standard fields, 
color equations to transform instrumental magnitudes into the
Johnson-Cousins system were computed for each CCD. 
The results are consistent with ESO\footnote{see ESO-WFI homepage} official
equations:
\begin{eqnarray}
\nonumber B_J&=&b_{99}+24.65+0.24\times (B_J-V_J) \\
\nonumber B_J&=&b_{123}+24.71+0.19\times (B_J-V_J)\\
\label{eq:col_curve} V_J&=&v_{89}+24.15-0.13\times (V_J-R_C) \\
\nonumber V_J&=&v_{89}+24.15-0.09\times (B_J-V_J) \\
\nonumber R_C&=&r_{162}+24.47\textrm{.}
\end{eqnarray} 
The $b_{99}$ or $b_{123}$ equation is adopted depending on
which filter was used (see Sect. \ref{sec:observations}).
\begin{figure}[!t]
\centering
\includegraphics[width=0.45\textwidth]{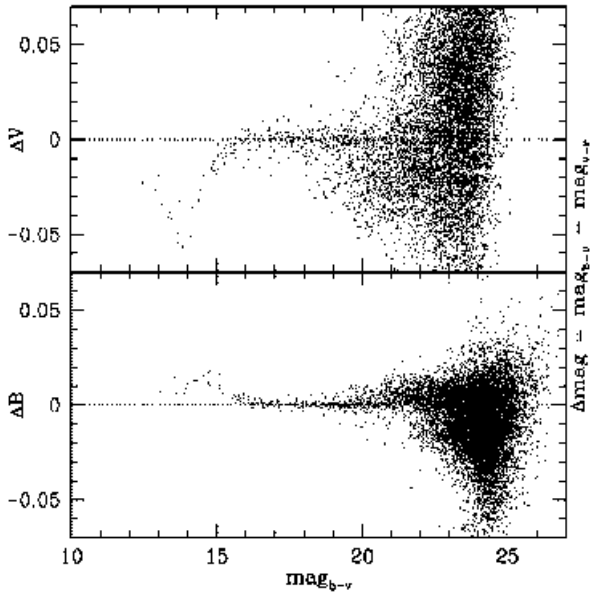}
\caption{Self-check of photometric calibration. The color equations (Eqs.
\ref{eq:col_curve}) built on three photometric bands, lead to two different
solutions, based on the $(B-V)$ and $(V-R)$ colors. 
The magnitudes based on the two calibrations turn out to be perfectly consistent
to each other.}
\label{fig:self_check}
\end{figure}
The catalog was calibrated to the
standard Johnson-Cousins photometric system ($B_J$, $V_J$, $R_C$), by solving
the above array iteratively.

Since 3 bands and 2 colors are available, two different
calibrations are possible, based on $(B-V)$ or $(V-R)$. 
Figure \ref{fig:self_check}
shows the difference in magnitude between the two different calibrations. 
When one color was not available, a spiral-like average of 1.0 was adopted.
Only B and V are shown in Fig. \ref{fig:self_check}, since no color 
term affects the R band filter. The two estimates are self-consistent; 
the associated magnitude uncertainty is similar or even smaller than that
estimated by SExtractor on background noise.

\section{Quality tests}\label{sec:quality}

In order to test data quality and survey performance, a series of 
tests have been performed, based on simulations,
aimed at determining:
\begin{itemize}
\item a semiempirical estimate of the effective depth reached and of
completeness;
\item a reliable estimate of magnitude uncertainty, overriding SExtractor's;
\item the reliability of SExtractor's stellarity flag.
\end{itemize}
To this end, we have added synthetic sources to our images, by using IRAF tasks
in the package {\tt artdata}.

\begin{figure}[!t]
\centering
\includegraphics[width=0.45\textwidth]{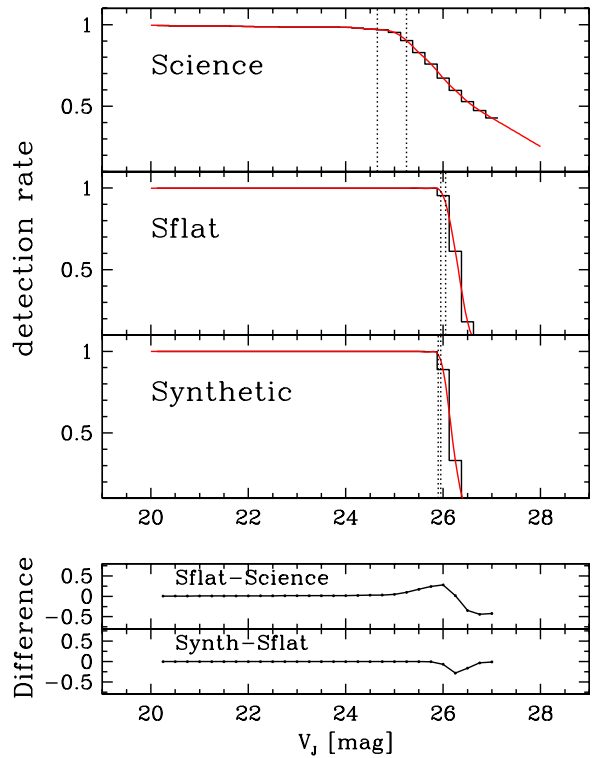}
\caption{Detection efficiency of synthetic sources in the V-band. Three cases
are studied: a real science frame already containing real objects, a {\em
ss-flat} image and a totally synthetic sky frame (see text for further
details). Histograms represent the real data, the continuous red line is a
spline interpolation. Vertical dotted lines set the 97\% and 90\% limits.
The bottom panel includes the difference in detection rate between the 
three cases. } 
\label{fig:compl_tests}
\end{figure}

The analysis comprises two steps.  First we have tested three different types of images
in order to understand the best way to reach our goal:
\renewcommand{\labelenumi}{\roman{enumi})}
\begin{enumerate}
\item the real image, containing all real sources; we chose WFI pointing number
2, because it represents the worst case, containing a very bright ($V\sim7.8$)
star;
\item an empty image obtained from the {\em non-smoothed} {\em ss-flat} frame, 
built from 30 individual
images belonging to the 6 ESIS central pointings;
\item a totally synthetic background image generated by IRAF, matching the {\em
r.m.s.} properties of the real image.
\end{enumerate}
\renewcommand{\labelenumi}{\arabic{enumi}.}
This test was run on the V band and using point-like objects only; results are
reported in Section \ref{sec:completeness}. The best choice turned out to be the
{\em ss-flat} frame, because it reproduces all the defects of real science
images, but is not affected by confusion problems.

A number of different simulated images per band were then produced, by
adding point-like, De Vaucouleurs or exponential disk objects onto the {\em
ss-flat}. 
In each case, a different image was produced per 0.25 magnitude bin, in the range
20-27 mag. A population of 2000 sources was added in each case, for a total of
$\sim$50000 objects on 14 different images per band.

Concerning De Vaucouleurs and disk profiles, intrinsic half-light radii between
2 and 10 $[$kpc$]$ and a Euclidean power law luminosity function were adopted.
The {\tt artdata} package assumes a redshift distribution that corresponds to
the apparent magnitude distribution. The standard cosmological dimming 
of flux and angular size are applied to each artificial galaxy. Finally the
synthetic objects are convolved with a gaussian kernel consistent with the real
seeing of observations and are added to the {\em ss-flat} frames.

The results described below are based on SExtractor's performance on this
simulated images.

\subsection{Detection rate}\label{sec:completeness}

Figure \ref{fig:compl_tests} summarizes the results of the preliminary test
performed on the V band images. The detection efficiency of SExtractor was defined
as $\epsilon=N_{SEx}/N_{in}$, i.e. the fraction of input synthetic
sources recovered by SExtractor.
The three top panels in Fig. \ref{fig:compl_tests} show the dependence of
$\epsilon$ on magnitude, measured on the real science final image, on
the {\em ss-flat} frame, and on a synthetic sky image.
The two bottom panels report the difference in $\epsilon$, between the various
cases.

The effects of confusion influence the detection of synthetic sources on the real
science frame, with respect to other cases, in two ways: $\epsilon$
decreases from unity at brighter magnitudes, because sources get lost 
in the halos of bright stars and very extended galaxies; 
$\epsilon$ decreases more slightly than in the other cases, because of
spurious detection due to superimposition onto real sources.

Concerning the {\em ss-flat} image, which was intentionally neither smoothed, nor
cleaned of bright halos, $\epsilon$ decreases less suddenly than on the
synthetic sky, because some spurious sources are detected in the halos.

We estimate the amount of spurious detections by running SExtractor on the
empty {\em ss-flat} (i.e. without any simulated object added), searching for
sources at the positions of those previously detected. No spurious sources were
detected down to V$\sim$25; at fainter fluxes about 1-2\% of sources are not
real, in the range 25-26 mag. Similar results were obtained in B and R.

Table \ref{tab:completeness} summarizes the resulting completeness values derived
for point-like, De Vaucouleurs and exponential-disk simulations on the {\em
ss-flat} images, in the three bands. Overall, a 90\% detection efficiency is
reached at mag$\sim$25.5, with some scatter, depending on the case.
For pointing number 6, consisting of 10 exposures only, this limit is $\sim0.5$
mag brighter.

\begin{table*}[!t]
\tiny
\centering
\begin{tabular}{c|ccc|ccc|ccc}
\hline
\hline
\multicolumn{10}{c}{Point-like sources}\\
\hline
& \multicolumn{3}{c|}{R} & \multicolumn{3}{c|}{V} & \multicolumn{3}{c}{B}\\
\hline
mag & $\sigma$ & s.i.q. & rate & $\sigma$ & s.i.q. & rate & $\sigma$ & s.i.q. & rate \\ 
\hline             
21	& 0.024	&	0.005	& 0.999	& 0.024	&	0.003	&	0.999	&	0.022	&	0.002	&	0.999\\
23	& 0.063	&	0.034	& 0.999	& 0.042	&	0.020	&	0.999	&	0.033	&	0.015	&	0.999\\
25	& 0.283	&	0.175	& 0.999	& 0.199	&	0.115	&	0.999	&	0.158	&	0.095	&	0.999\\
26	& 0.478	&	0.266	& 0.064	& 0.291	&	0.178	&	0.953	&	0.282	&	0.172	&	0.980\\
\hline               
\multicolumn{10}{c}{}\\
\multicolumn{10}{c}{}\\
\hline               
\multicolumn{10}{c}{Exponential-disk sources}\\
\hline               
& \multicolumn{3}{c|}{R} & \multicolumn{3}{c|}{V} & \multicolumn{3}{c}{B}\\
\hline
mag & $\sigma$ & s.i.q. & rate & $\sigma$ & s.i.q. & rate & $\sigma$ & s.i.q. & rate \\ 
\hline             
21	&	0.060	&	0.027	&	0.996	&	0.074	&	0.028	&	0.994	&	0.039	&	0.010	&	0.993\\
23	&	0.142	&	0.084	&	0.994	&	0.371	&	0.166	&	0.992	&	0.089	&	0.035	&	0.989\\
25	&	0.470	&	0.298	&	0.931	&	1.042	&	0.602	&	0.983	&	0.254	&	0.147	&	0.981\\
26	&	0.858	&	0.275	&	0.018	&	1.195	&	0.678	&	0.522	&	0.456	&	0.256	&	0.693\\
\hline               
\multicolumn{10}{c}{}\\
\multicolumn{10}{c}{}\\
\hline               
\multicolumn{10}{c}{De Vaucouleurs sources}\\
\hline               
& \multicolumn{3}{c|}{R} & \multicolumn{3}{c|}{V} & \multicolumn{3}{c}{B}\\
\hline
mag & $\sigma$ & s.i.q. & rate & $\sigma$ & s.i.q. & rate & $\sigma$ & s.i.q. & rate \\ 
\hline             
21	& 	0.064	&	0.033	&	0.998	&	0.084	&	0.038	&	0.997	&	0.035	&	0.014	&	0.997\\
23	&	0.182	&	0.102	&	0.993	&	0.446	&	0.210	&	0.994	&	0.082	&	0.042	&	0.990\\
25	&	0.512	&	0.291	&	0.811	&	1.148	&	0.658	&	0.976	&	0.322	&	0.181	&	0.974\\
26	&	0.829	&	0.327	&	0.009	&	1.218	&	0.701	&	0.268	&	0.455	&	0.279	&	0.443\\
\hline
\end{tabular}
\footnotesize
\caption{Summary of quality simulations on the {\em ss-flat} image. Three cases
are reported, consisting of point-like, De Vaucouleurs and exponential-disk
synthetic objects. For each band, magnitude standard deviation and 
semi-inter-quartile (s.i.q.) are listed, as well as extraction efficiency.}
\normalsize
\label{tab:completeness}
\end{table*}

\subsection{Magnitude uncertainty}\label{sec:phot_err}

\begin{figure}[!t]
\centering
\includegraphics[width=0.45\textwidth]{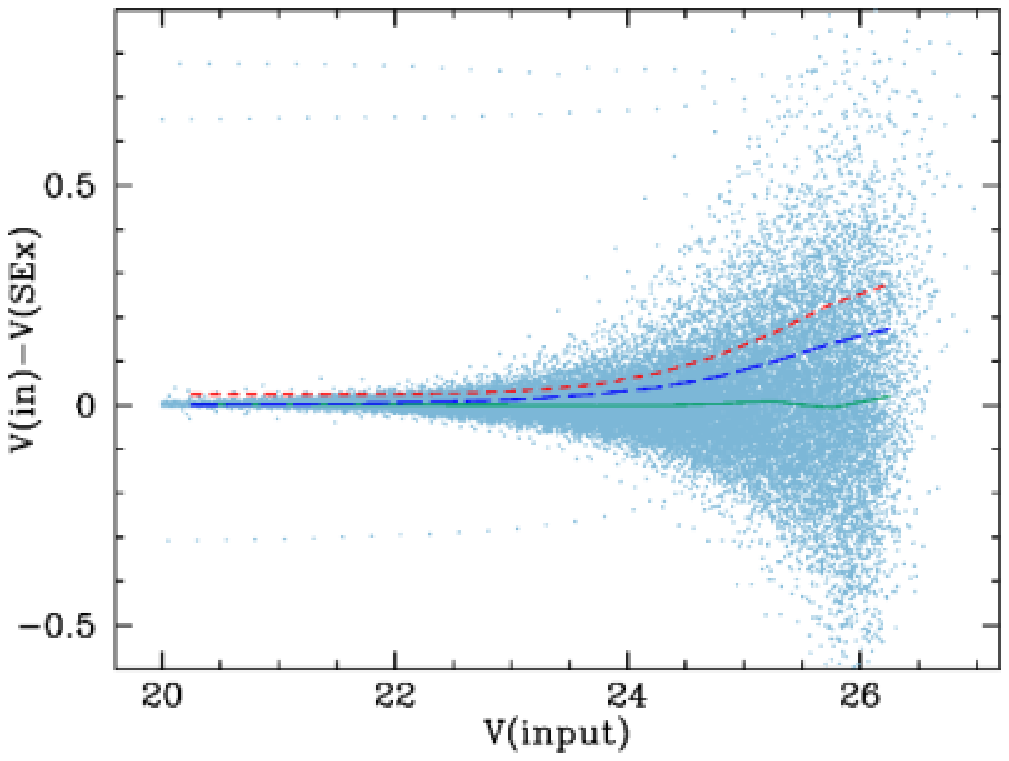}
\rotatebox{-90}{\includegraphics[height=0.45\textwidth]{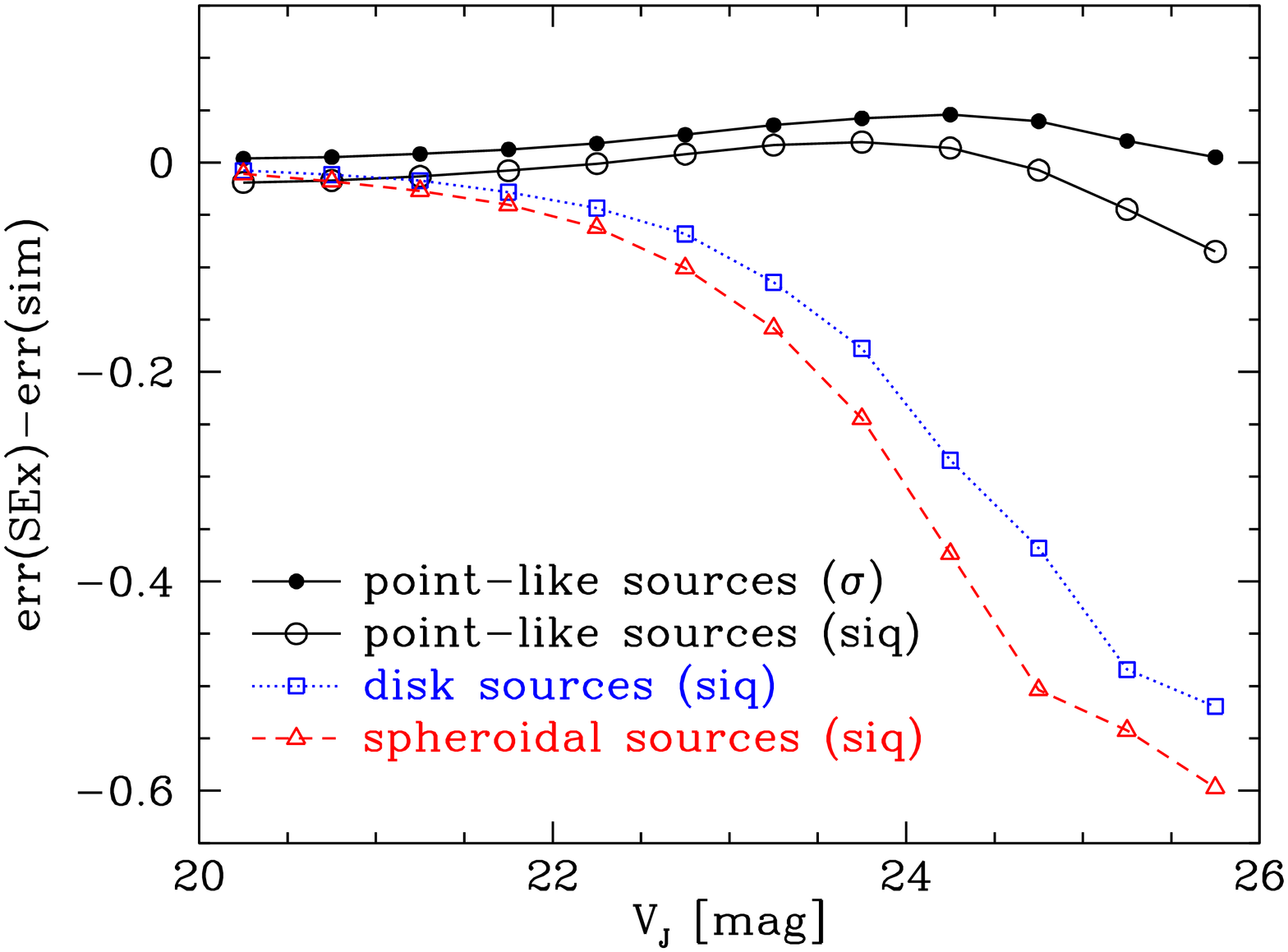}}
\caption{{\em Top panel}: V-band {\em ss-flat} simulation. Difference between 
the input magnitude
of point-like synthetic sources and the value measured by SExtractor, as a
function of input mag. The solid line traces the median, the short-dashed 
standard deviation, and the long-dashed semi-inter-quartile (s.i.q.).
{\em Bottom panel}: comparison between the r.m.s. (or s.i.q.) computed 
on simulated data and 
the median magnitude uncertainty measured by SExtractor on real data, 
in a given magnitude bin. Solid, dotted and dashed lines represent point-like, 
disk and spheroidal sources, respectively.}
\label{fig:mag_unc}
\end{figure}

\begin{figure*}[!t]
\centering
\includegraphics[height=0.37\textwidth]{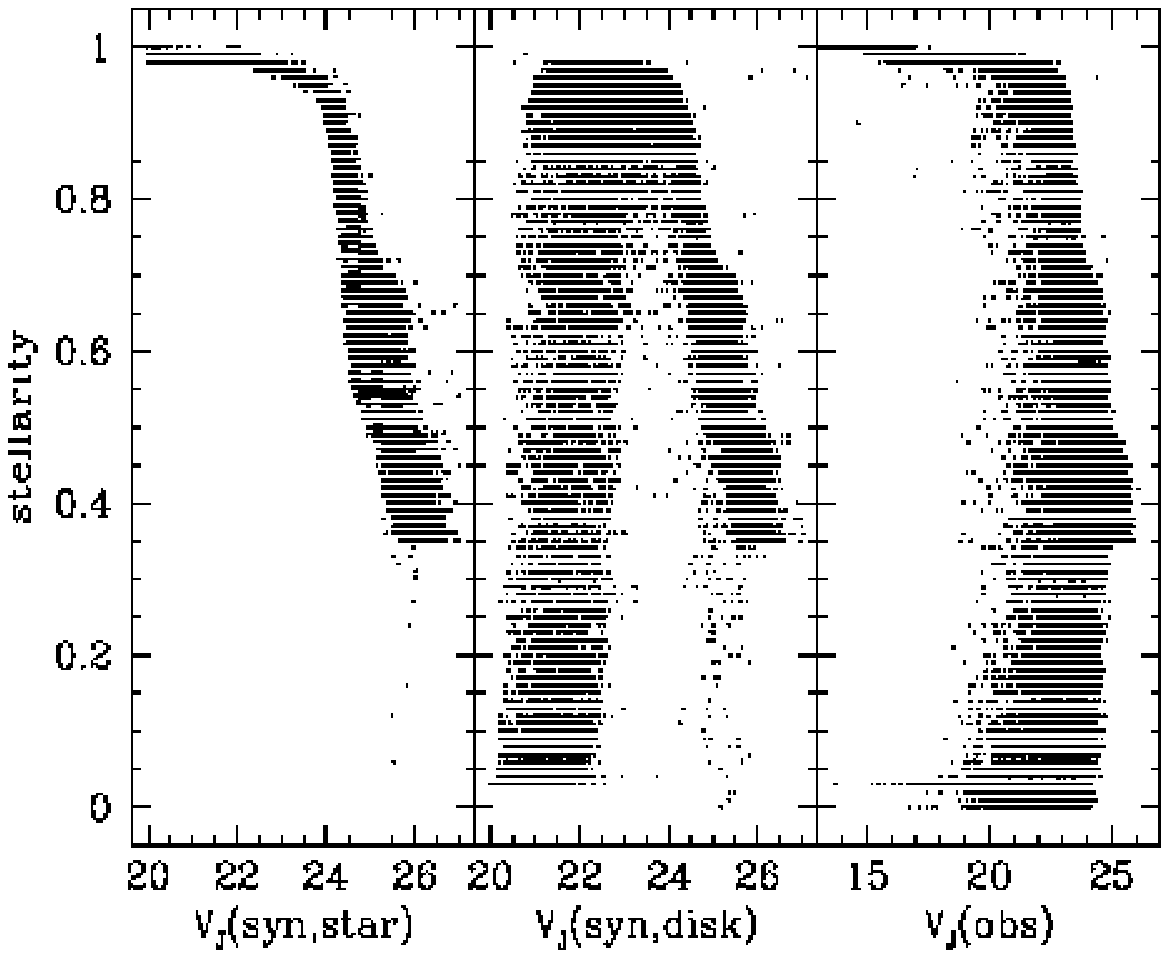}
\includegraphics[height=0.37\textwidth]{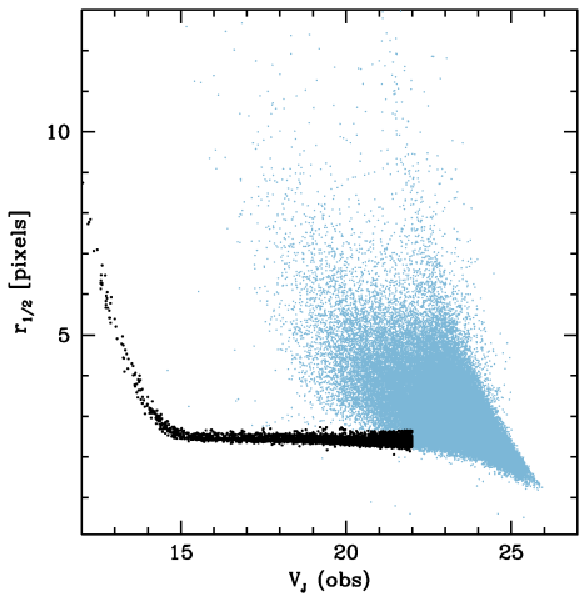}
\caption{{\em Left}: dependence of the stellarity index on $V_J$ magnitude, 
for simulated point-like sources, simulated exponential disks and real data. 
The horizontal dotted line sets the CLASS\_STAR$=$0.95 limit.
{\em Right}: half-flux radius as a function of $V_J$ magnitude. The dark dots
define the locus of point-like sources.}
\label{fig:stellarity}
\end{figure*}

An estimate of magnitude uncertainties is obtained by
comparing the input simulated fluxes and the measured values.
The top panel in Fig. \ref{fig:mag_unc} shows the case of point-like sources added to the V
band {\em ss-flat} frame: the difference between input and extracted magnitudes
is plotted as a function of input magnitude. The solid line represents the
median difference, the short-dashed line is the standard deviation from the
median value and the long-dashed line traces the
semi-inter-quartile\footnote{defined as $(q_3 - q_1)/2$, where $q_1$ and $q_3$
are the first and third quartiles. The first quartile is the number below which 25\%
of the data are found and the third
quartile is the value above which 25\% of the data are found.}.

The bottom panel in Fig. \ref{fig:mag_unc} compares the 
standard deviation and semi-inter-quartile of simulations (see top panel)
to the median magnitude uncertainties measured by SExtractor on real data.
The latter is computed as the median of uncertainties for all real sources 
in a given magnitude bin. 
In the case of point-like sources, SExtractor's errors are compared both to the 
standard deviation from the median (filled circled) and to the
semi-inter-quartile (open circles). The synthetic $\sigma$ is typically driven
by outliers (such as unresolved double stars), while the semi-inter-quartile
width of the distribution provides a fairer comparison to SExtractor's
uncertainties.

The estimate of magnitude errors for non-pointlike objects is compared to the
simulated ones for disk and spheroidal galaxies (open squares and triangles
respectively). The discrepancy between the two estimates is larger in this case, 
because of non-shot-noise effects like failure in detecting the faint wings
of galaxy profiles or size dimming for the faintest sources.

Table \ref{tab:completeness} summarizes the results for all the cases
studied. Typical uncertainties at mag$=21,\ 25$ are $\sim 0.03,\ 0.3$ in
the B band, $\sim 0.05,\ 0.5$ in V and R.
An uncertainty of 0.05 mag at 21 corresponds to one of 5\% in flux, while
0.5 at 25 corresponds to 60\% in flux.

\subsection{Stellarity}\label{sec:stellarity}

Finally, the simulations described above are also useful to test the reliability
of SExtractor's {\em stellarity} flag. In combination with real data, 
it is possible to evaluate down to which magnitude the CLASS\_STAR flag
represents a realistic indication of the source profile.

The left plots in Fig. \ref{fig:stellarity} show the distribution of CLASS\_STAR
as function of magnitude, for the simulated point-like sources (left panel), 
simulated exponential disks (central panel), and for the real data 
(right panel). 

\begin{figure*}[!t]
\centering
\includegraphics[width=0.32\textwidth]{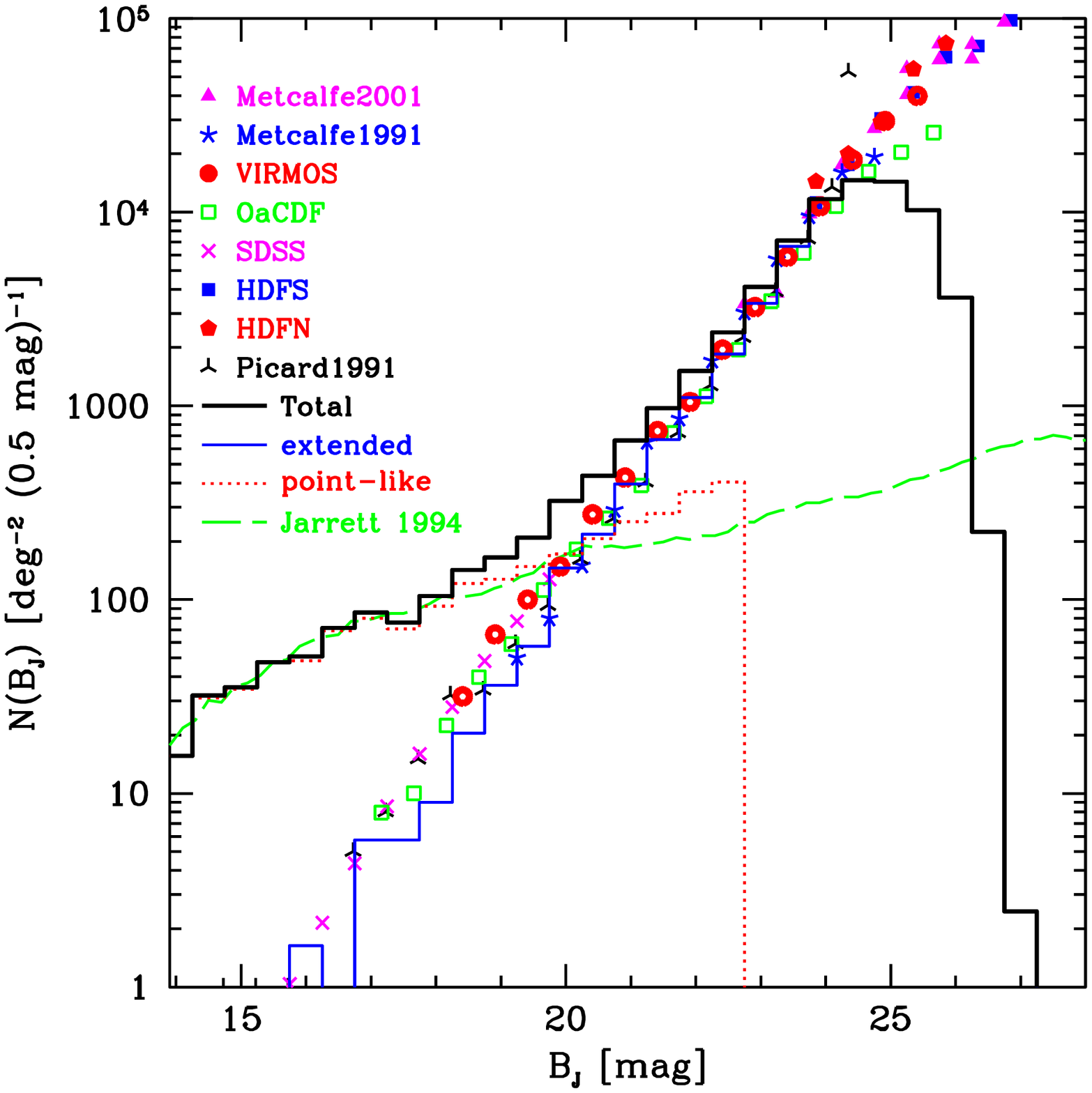}
\includegraphics[width=0.32\textwidth]{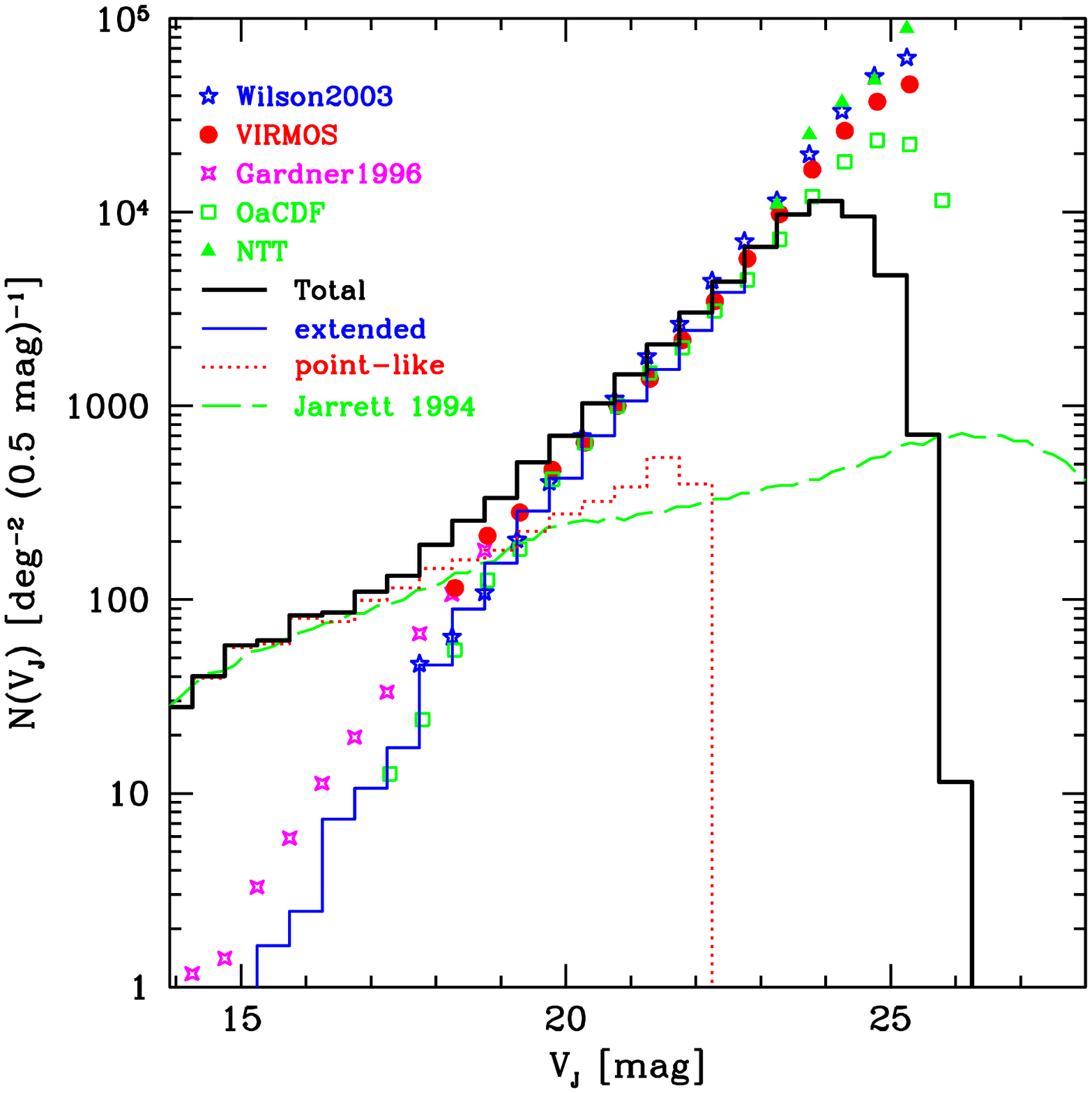}
\includegraphics[width=0.32\textwidth]{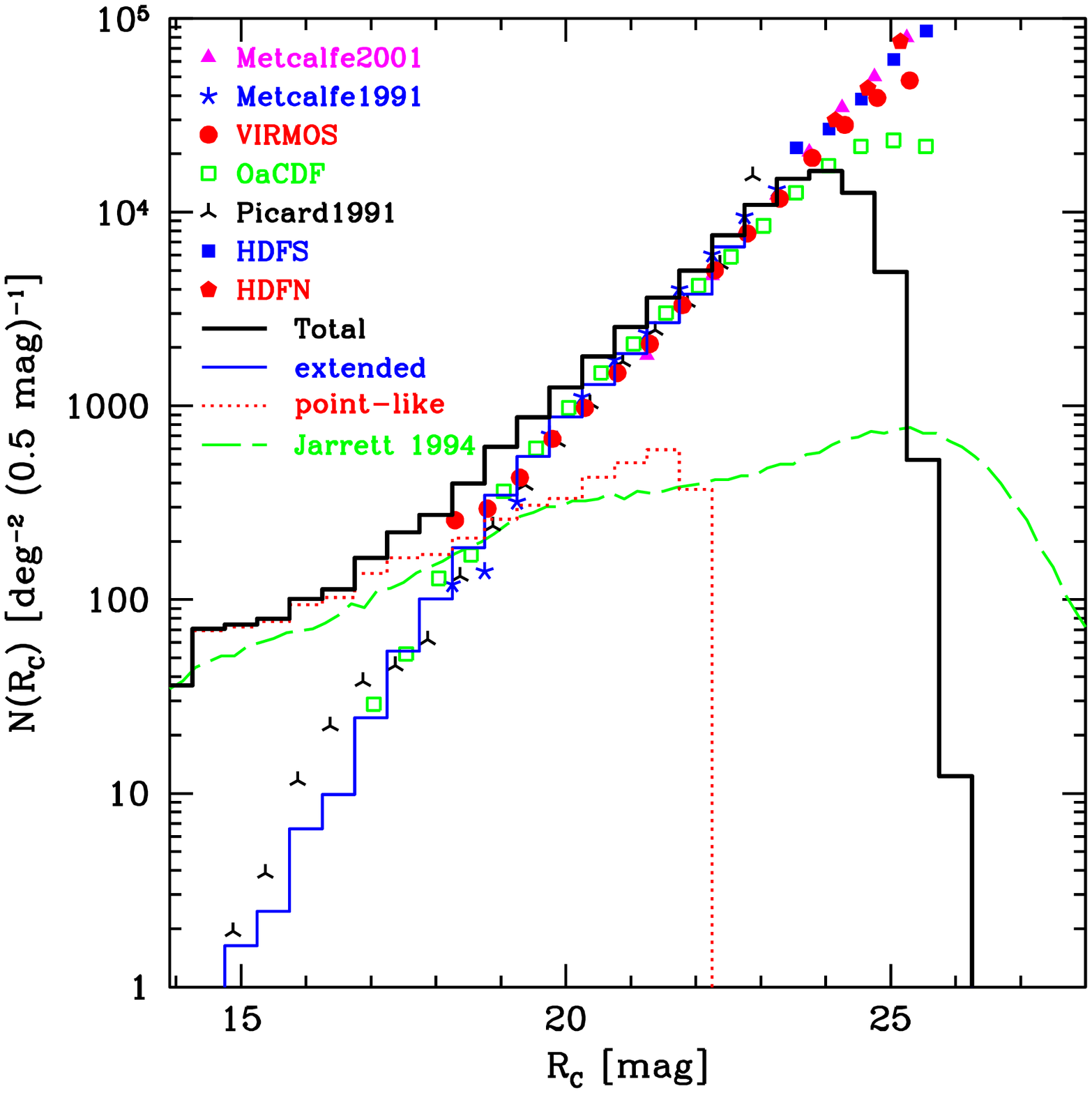}
\caption{Observed ESIS number counts (histograms), compared to literature data
(dots) and \citet{jarrett1994} Milky Way model (long-dashed line). 
The ESIS data are split into extended (this solid histograms) and point-like (dotted) components. 
The bold solid histograms represents the total ESIS counts.}
\label{fig:counts}
\end{figure*}

The plots suggest that a CLASS\_STAR$>$0.95 is a good representation of
point-like sources up to V$\sim23$. Note that effectively no point-like sources are
detected above this threshold, at fainter fluxes. Nevertheless, a significant
contamination by galaxies is expected at fluxes fainter that V$\sim$21, 
as confirmed by Fig.
\ref{fig:counts} on real data (see also Section \ref{sec:stars}).
Similar results are obtained in the other two bands, with slightly fainter
magnitude limits.

An intrinsically wide scatter in CLASS\_STAR is typical of galaxies. 
Moreover these tend
to behave like point-like sources at faint fluxes, because
the profiles of small objects are dominated by seeing.

A safer way of extracting point-like sources relies on the 
half-light radius, $r_{1/2}$, enclosing 50\% of the object's total flux.
As for point-like sources $r_{1/2}$ depends only on the
image seeing, when plotted against magnitude it defines a tight 
``stellar'' locus (see the diagram on the right side of Fig. \ref{fig:stellarity}). 
Although still contaminated by QSOs, this method provides a good 
representation of point-like objects down to V$\sim22$.

\section{Source counts}\label{sec:counts}

The magnitude distribution of galaxies, per unit sky area, provides strong
constraints on their evolutionary history \citep{ellis1997,metcalfe1995}.
In monolithic-collapse models \citep{eggen1962}, the bulk of 
stars formed in early-type galaxies at very high redshift, and galaxies evolve
passively to present day; while subsequent merging and star formation are
limited. 
In a hierarchical merger scenario \citep{kauffmann1998}, the bulk of the stars
form in disk-like galaxies that later merge to become early-type galaxies.
The luminosity and density of galaxies strongly evolve with cosmic time due to
mergers and related starbursting events.

We build ESIS counts excluding WFI pointing number 6, hence including sources within a
1.22 $[$deg$^2]$ area. Figure \ref{fig:counts} shows the data: the solid thick
histogram represents the total observed number counts, the dotted histogram the
counts for point-like sources (see Sect. \ref{sec:stellarity}), and the thin 
solid histogram the counts for extended objects.

\subsection{Star-galaxy separation}\label{sec:stars}

In order to compare to galaxy evolutionary models, stars must be removed 
from number counts.
On the basis of the analysis presented in Sect. \ref{sec:stellarity}, 
the half-light radius $r_{1/2}$ provides a good representation 
of point-like sources, but this method is limited to magnitudes brighter 
than $\sim22$. 
Moreover, it is worth to note that quasars are intrinsically point-like 
sources and cannot be a priori distinguished from stars, on the basis 
of BVR colors only. 

The long-dashed line in Figure \ref{fig:counts} represents the expected
magnitude distribution for stars, based on the Milky Way model by 
\citet{jarrett1992} and \citet{jarrett1994}. These authors adopt a three
component model, including a halo, a thin disk and a thick disk stellar
population. This model reproduces the observed
star counts in different galactic lines of sight, as well as SDSS data.

The Milky Way model is consistent with ESIS
data at bright fluxes, in the B and V bands, while it seems to underpredict (by
about 20-25\%)
star counts in the R band at magnitudes $<19$. Between magnitudes 21 and 22,
observed pointlike counts still show some excess with respect to
Jarrett model. This may be caused by residual contamination from faint
unresolved galaxies or by a quasar component.

In what follows (e.g. Figure \ref{fig:counts_model}), two corrections will be
applied to number counts for stars: one based on the \citet{jarrett1994}
model, the other using SExtractor's identification. Overall the latter
correction is more solid than the former at bright fluxes, while the former is
most suitable for faint magnitudes.   In any case it should be noted that the
stellar contribution to the
counts at the faint end (less than 3\%) is almost negligible, when comparing
data and galaxy evolutionary models, while it dominates the bright counts.

\subsection{Comparison to literature data}

Figure \ref{fig:counts} also compares ESIS galaxy counts to a collection of
literature data. In particular, HDFN \citep{williams1996}, HDFS
\citep{metcalfe2001}, SDSS \citep{yasuda2001}, VIRMOS deep survey
\citep{mccracken2003}, OaCDF \citep{alcala2004}, NTT SUSI Deep Field
\citep{arnouts1999}. Additional data are from \citet{metcalfe1991,metcalfe2001},
\citet{picard1991}, \citet{wilson2003} and \citet{gardner1996}. All magnitudes
have been transformed to the Johnson-Cousins photometric system, adopting the
appropriate relations (usually reported by authors).

ESIS number counts are generally consistent with the results from other surveys, in the
common flux range. At bright magnitudes, some
differences arise between our counts for extended sources (which can be
considered galaxies for mag$<$18). This is particularly true in the V band, the 
\citet{gardner1996} data being significantly higher than ESIS', possibly due to
a different choice in accounting for stars, or to cosmic variance.

\subsection{Galaxy evolutionary models}\label{sec:counts_mod}

Differential galaxy number counts can be described as the integral over cosmic
time of the galaxy luminosity function (LF), in a specific flux bin
\citep[e.g.][]{franceschini2001,guiderdoni1991}. The shape
and normalization of the LF is assumed constant in non-evolutionary models;
galaxy luminosity evolves by means of galaxy ages, in {\em pure luminosity
evolution} models (PLE); finally in complete evolutionary models, the LF changes
as a function of redshift, for example due to enhanced star formation
activity, triggered by encounters  and mergers between galaxies.

In the recent years hierarchical clustering models have gained 
increasing popularity in the description of galaxy formation and
evolution. Primordial density perturbations are amplified by
gravitation and collapse to form almost virialized structures called
dark matter (DM) haloes. In the {\em cold dark matter}
(CDM) scenario \citep{peebles1982,blumenthal1984}, smaller
haloes form first, subsequently merging into bigger haloes.
Gas cools down in the potential wells of the haloes,
leading to the formation of stars and galaxies. The spectrophotometric 
evolution of stellar populations, combined with the history
of stellar formation of galaxies, finally produce 
their observable properties: spectra, colors, etc.
Semi-analytic modern models combine these and others ingredients 
and follow the various processes transforming the 
primordial power spectrum of density fluctuations into 
the spectral energy distributions of stellar populations and galaxies.
\citep[e.g.][]{hatton2003,guiderdoni1998}.

\begin{figure*}[!t]
\centering
\includegraphics[width=0.4\textwidth]{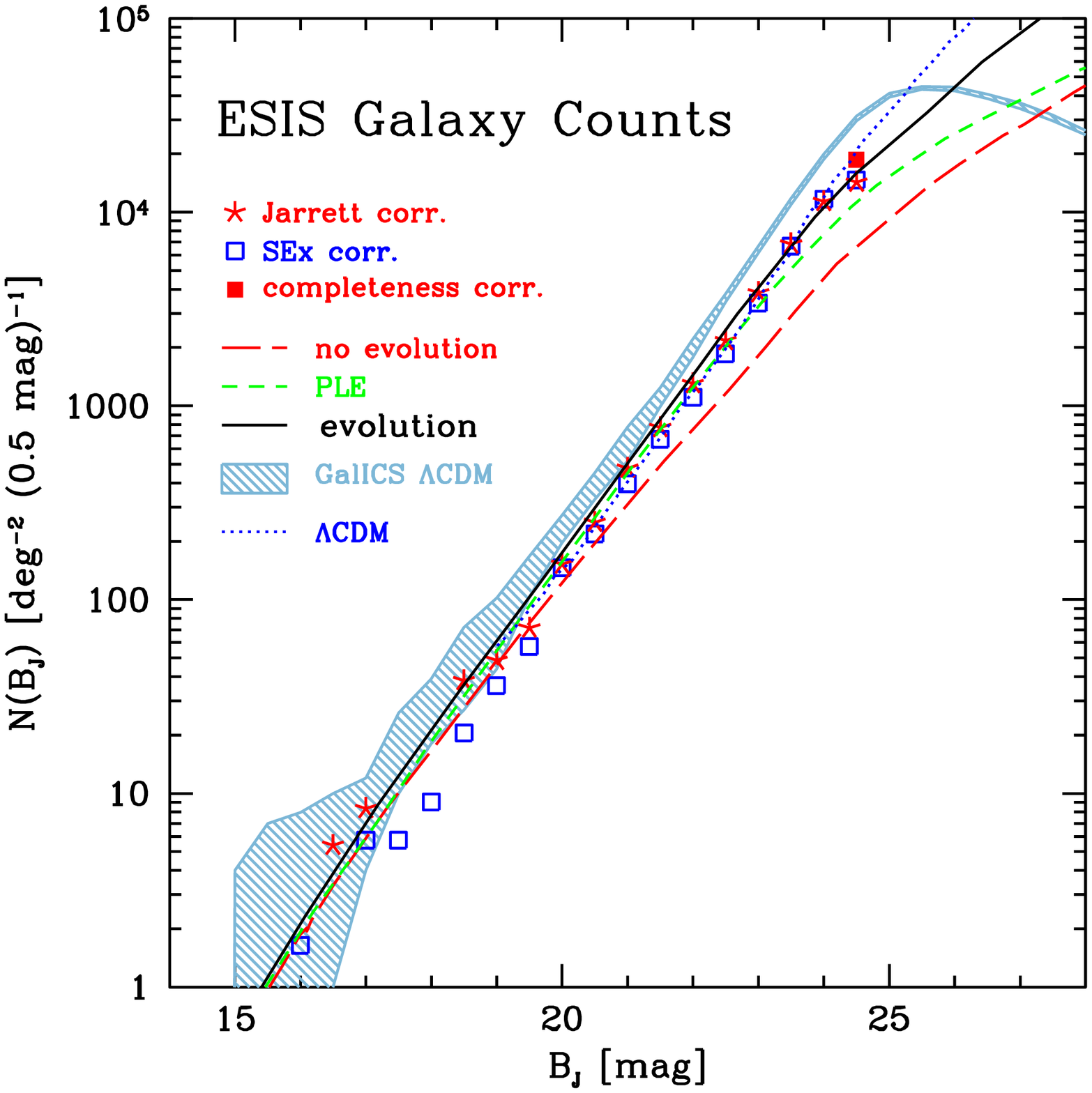}
\includegraphics[width=0.4\textwidth]{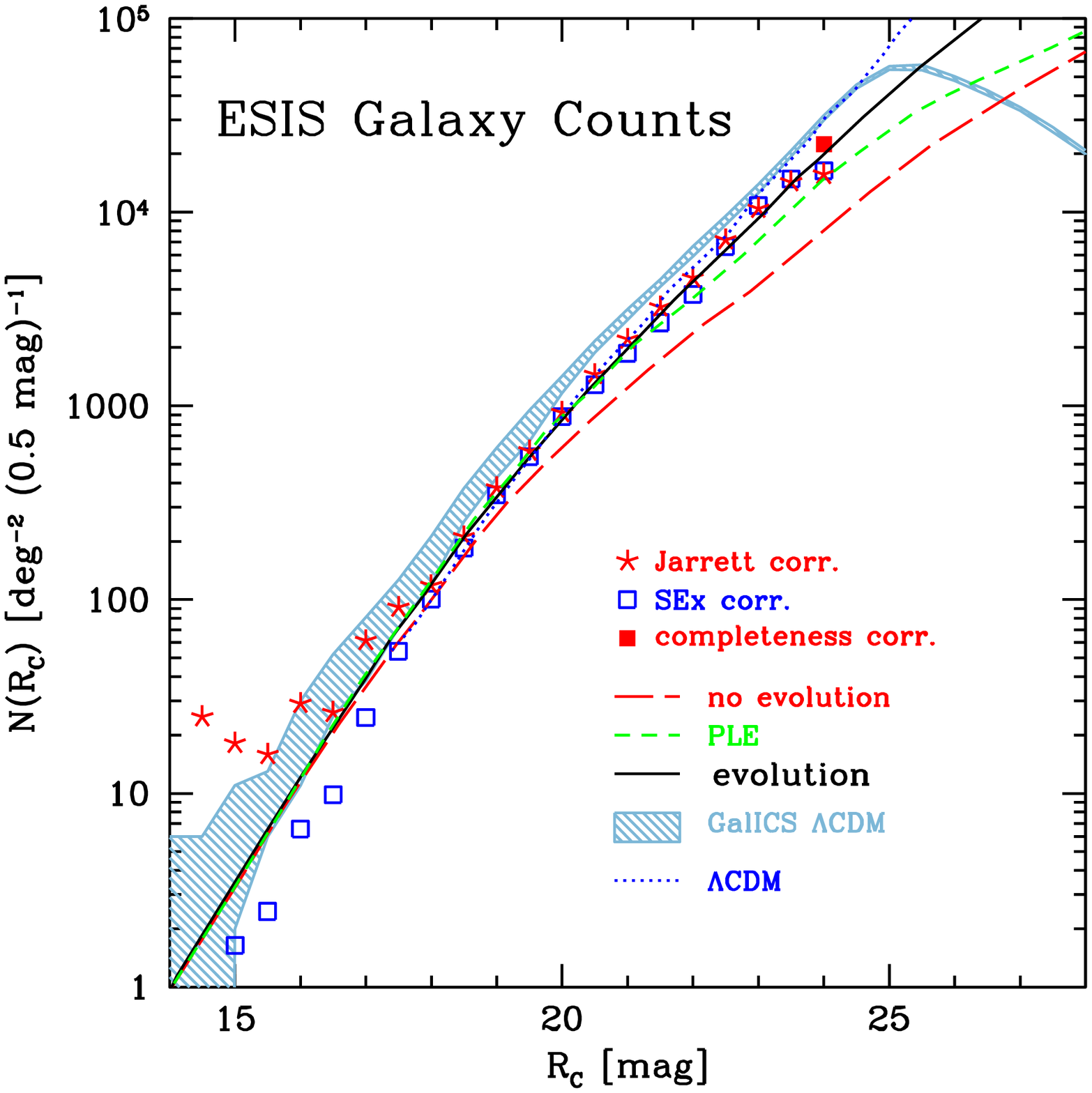}
\caption{Comparison between ESIS galaxy counts and prediction of evolutionary
scenarios. Open squares are number counts for extended sources (based on
the half-flux radius analysis). Asterisks represent total galaxy counts corrected 
for stars contribution with the
\citet{jarrett1994} Milky Way model. Filled squares are data corrected for
completeness (at the faintest fluxes). 
Models belong to \citet{metcalfe2001}, \citet{nagashima2002} and the GalICS team
\citep[shaded area,][]{hatton2003}.} 
\label{fig:counts_model}
\end{figure*}

Figure \ref{fig:counts_model} compares ESIS galaxy counts to the four different
kinds of models. Galaxy counts are obtained from total number counts in two
ways: by subtracting point-like identifications (open squares) and by using the
\citet{jarrett1994} model (stars). Poisson errors are smaller than the symbols, for
mag$>$18.

\begin{table}[!t]
\centering
\footnotesize
\begin{tabular}{c|cc|cc|cc}
\hline
\hline
   & \multicolumn{2}{c|}{B$_J$} & \multicolumn{2}{c|}{V$_J$} & \multicolumn{2}{c}{R$_C$}\\
\hline
mag &  $\log N $ & $\log N$ & $\log N$ & $\log N$ & $\log N$ & $\log N$ \\
    &  (Ext.) &  (Tot.) & (Ext.)  & (Tot.) & (Ext.) & (Tot.)\\        
\hline
16.0   &   0.215    &	1.713	 &   0.391   &    1.926 &  0.817   &   2.011  \\
16.5   &   0.215    &	1.848	 &   0.759   &    1.922 &  0.993   &   2.044  \\
17.0   &   0.692    &	1.935	 &    1.090  &    2.041 &   1.432  &   2.228  \\
17.5   &   0.759    &	1.900	 &    1.294  &    2.128 &   1.740  &   2.348  \\
18.0   &   0.993    &	2.014	 &    1.654  &    2.277 &   2.004  &   2.443  \\
18.5   &   1.345    &	2.152	 &    1.943  &    2.406 &   2.272  &	 2.600\\
19.0   &   1.567    &	2.228	 &    2.171  &    2.524 &   2.544  &   2.791  \\
19.5   &   1.752    &	2.329	 &    2.460  &    2.704 &   2.743  &   2.945  \\
20.0   &   2.164    &	2.508	 &    2.612  &    2.838 &   2.949  &   3.099  \\
20.5   &   2.359    &	2.652	 &    2.849  &    3.014 &   3.114  &	3.260 \\
21.0   &   2.596    &	2.820	 &    3.023  &    3.159 &   3.275  &   3.412  \\
21.5   &   2.836    &	2.996	 &    3.184  &    3.312 &   3.433  &   3.562  \\
22.0   &   3.050    &	 3.184   &    3.386  &    3.479 &   3.582  &   3.703  \\
22.5   &   3.275    &	 3.386   &    3.584  &    3.641 &   3.828  &   3.887  \\
23.0   &   3.540    &	 3.621   &    3.815  &    3.820 &   4.043  &   4.047  \\
23.5   &   3.834    &	 3.864   &    3.997  &    3.997 &   4.183  &   4.183  \\
24.0   &   4.077    &	 4.078   &    4.120  &    4.120 &   4.350  &   4.350  \\
24.5   &   4.271    &	 4.271   &    4.224  &    4.224 &   --     &   --     \\
\hline
\end{tabular}
\caption{ESIS BVR number counts $[$deg$^{-2}\ ($0.5 mag$)^{-1}]$.
Completeness corrections has been applied, based on simulations of galaxy-like
sources (see Sect. \ref{sec:quality}). Total and extended source counts
are provided.} 
\normalsize
\label{tab:counts}
\end{table}

Completeness correction is attempted at the faintest
fluxes, on the basis of the detection rate analysis carried out in Section
\ref{sec:quality}. Only one magnitude bin can be fully recovered, reaching
B,V$\sim$24.5 and R$\sim24$. Table \ref{tab:counts} lists the ESIS counts.

Describing the data as a power law 
\begin{equation}
\log N\propto \gamma\times \textrm{mag,}
\end{equation}
we find
$\gamma_B=0.48$ in the range $B=19-24$, similarly to \citet{ellis1997} and
\citet{metcalfe2001}. The V-band data can be fitted with a double power law, having
an elbow at $V\sim20$: $\gamma_V=0.56$ between 16 and 20, and $\gamma_V=0.37$
for $V=20-24$. Finally, concerning R band data, the galaxy counts are well
reproduced by $\gamma_R=0.58$ in the range $16-20$ mag and $\gamma_B=0.32$ for
$R=20-24$, consistent with \citet{metcalfe2001}.
Models are compared to data only for B and R band, because --- in the literature ---
the R band is usually preferred to the V band. 

In Fig. \ref{fig:counts_model}, the long-dashed line represents the
no-evolution model, by \citet{metcalfe2001}, obtained for a $q_0=0.5$ 
cosmology. These authors model the galaxy luminosity function (LF) as
a Schechter function, with $\alpha=-0.7$ for E/S0/Sab galaxies, $\alpha=-1.1$ 
for Sbc spirals and $\alpha=-1.5$ for bluer Scd/Sdm ones. 
In the B-band the model was normalized to $B\sim18$, instead of 15
\citep[see][ for a discussion in support of this
choice]{metcalfe1991,metcalfe1995,metcalfe2001} and reproduces reasonably well the 
observed counts down to $B\sim21$. In the R band the consistency with data holds only
to R$\sim19.5$. At fainter fluxes the model quickly diverges from
the ESIS counts. 

The short-dashed line in Fig. \ref{fig:counts_model} is a PLE model 
\citep[][]{metcalfe2001}, built assuming that the evolution of the
different morphological types is governed by their star formation history. 
Exponentially decaying star formation rates are adopted \citep{bruzual1993}, using
$\tau=2.5$ Gyr for E/S0 and Sab galaxies and $\tau=9$ Gyr for other spirals
types.
Present day galaxy ages are $\sim12.7$ Gyr, implying a formation redshift of
$z_f=9.9$.
This model is in good agreement with ESIS data down to B,R$\sim$23, after which 
the accordance fails.

\citet{metcalfe2001} introduced a population of dwarf ellipticals to increase the
model-predicted counts at the faintest fluxes \citep[down to B,R$\simeq28$, see also
][]{ellis1997} and match
observations. Alternative solutions lead to equally good fits, for example 
merging evolutionary models.

The solid line in Fig. \ref{fig:counts_model} represents an evolutionary
model, with a LF slope of $\alpha=-1.75$ for blue Scd/Sdm galaxies at redshift 
$z>1$, steeper than the local $\alpha=-1.5$ \citep{metcalfe2001}. 
This model provides a good fit to ESIS data over the entire magnitude range 
considered.

The GalICS 
\citep[Galaxies In Cosmological Simulations,][]{hatton2003}
project provides a hybrid model for hierarchical
galaxy formation studies, combining large cosmological $N$-body simulations
with semi-analytic recipes to describe the properties of galaxies within
dark matter haloes.
We have retrieved mock catalogs produced with the GalICS model, 
for a total of 10 different fields of 1 $[$deg$^2]$, in order to 
take into account the cosmic variance of the mock catalogs.
 
GalICS simulations assume a $\Lambda$CDM
flat Universe ($\Omega_m=0.333$, $\Omega_\Lambda = 0.667$, $h=0.667$, 
$\sigma_8 = 0.88$). 
Each realization consists of a $100~h^{-1}$ $[$Mpc$]$ cube,
containing $256^3$ particles of mass
$8.272\cdot10^9\ M_\odot$. The spatial resolution of the simulation 
is 29.29 $[$kpc$]$; the evolution of the Dark Matter density field 
is followed from $z=35.59$ to $z=0$, through 70 different snapshots.

Dark matter halos containing more than 20 particles (i.e. with
$M\ge1.66\cdot10^{11}\ M_\odot$) are identified and their merging history trees 
are then computed. Baryons
are evolved within these halo merging history trees, following a set of
semi-analytic prescriptions \citep[the Mock Map Facility, MoMaF,][]{blaizot2005}
that account for (among other effects) heating and cooling 
of the gas within halos, star formation and supernovae feedback on the
environment, evolution of stellar populations, metal enrichment, disk
instabilities, tidal stripping, and formation of
spheroids through galaxy mergers.

\begin{figure*}[!t]
\centering
\includegraphics[width=0.48\textwidth]{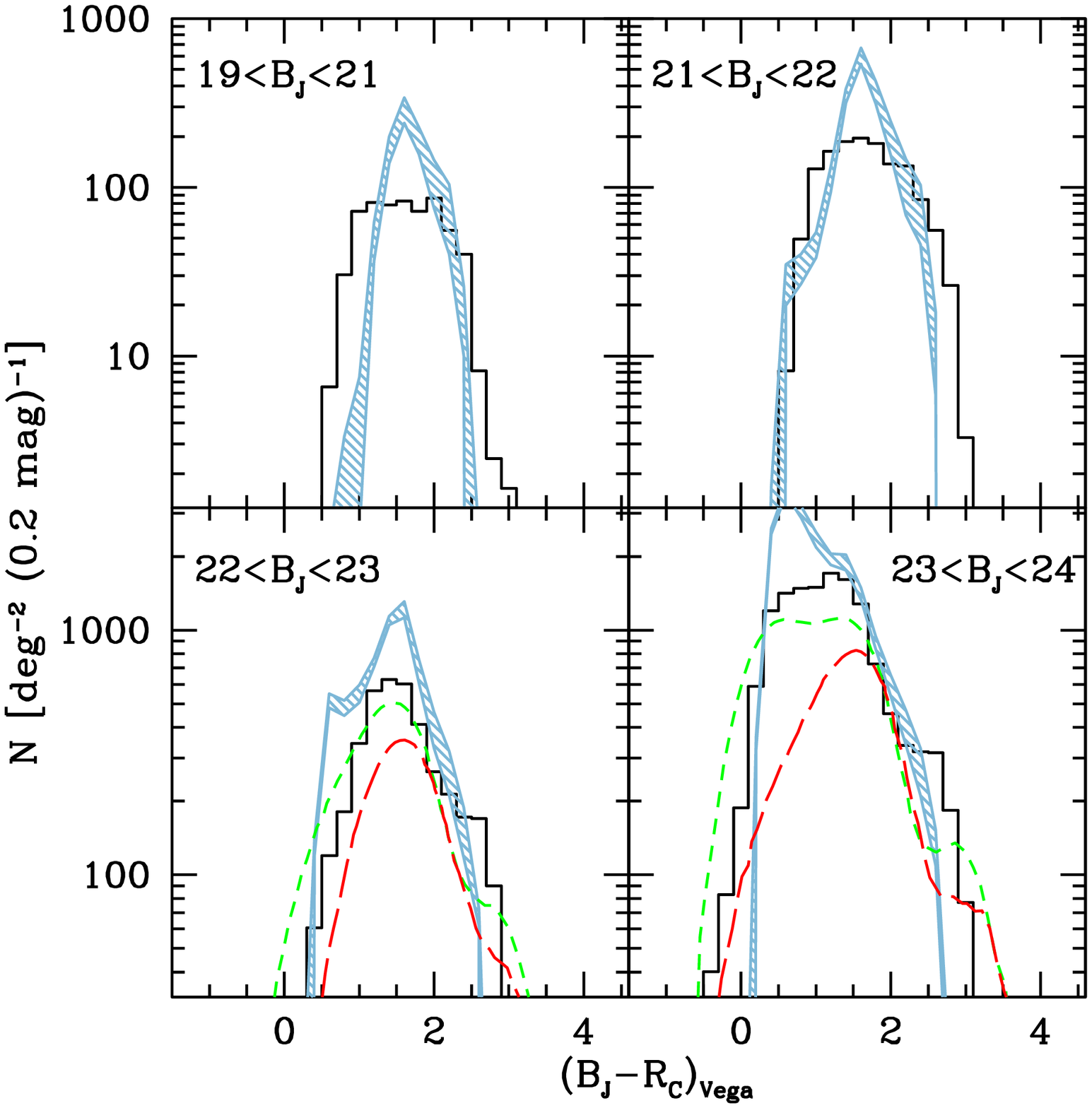}
\includegraphics[width=0.473\textwidth]{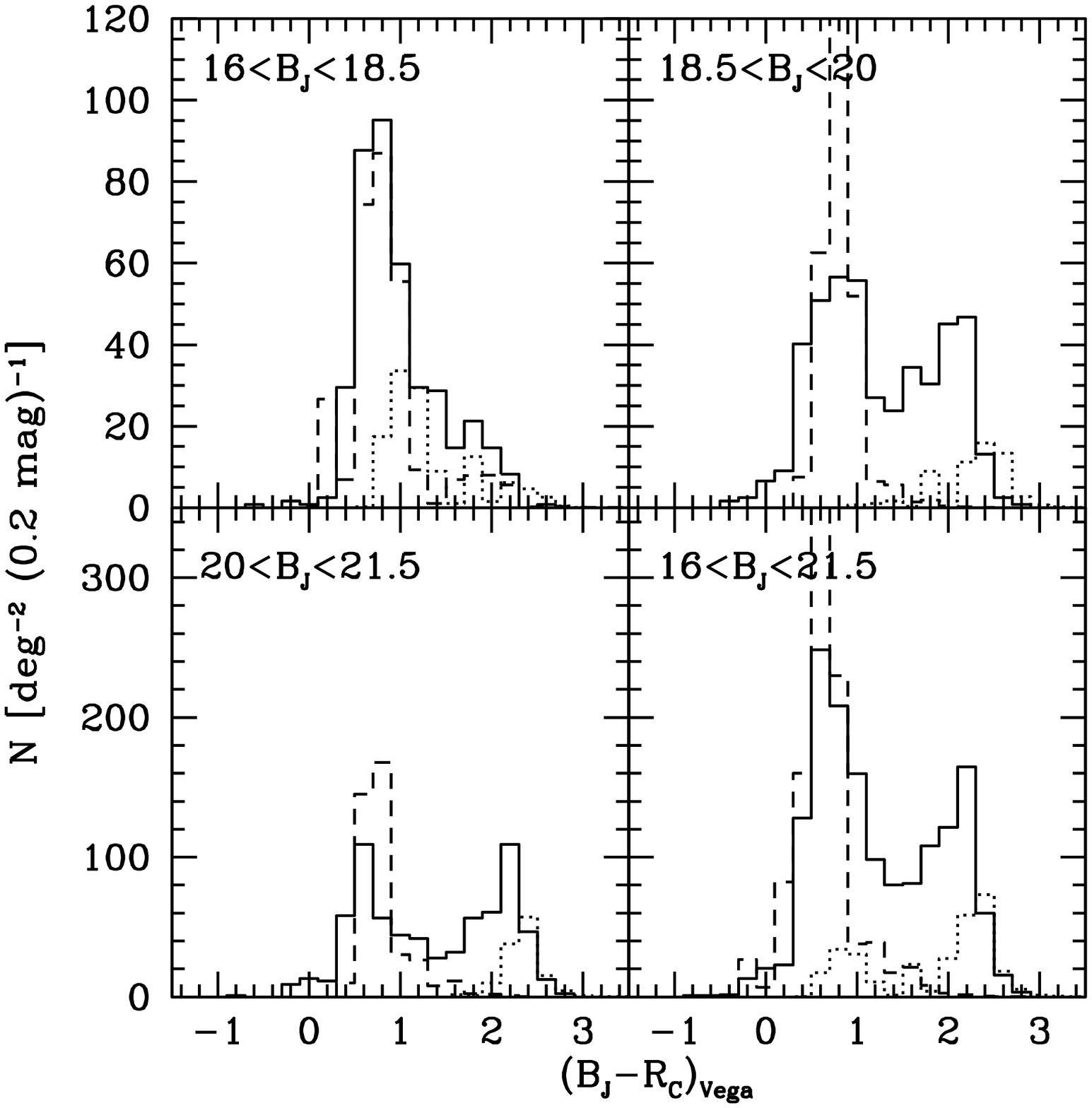}
\caption{Observed $B_J-R_C$ color distribution. {\em Left}: extended sources.
Overplotted are: a no-evolution \citep[long-dashed
line,][]{roche1996}, a PLE  \citep[short-dashed,][]{roche1996} and GalICS
$\Lambda$-CDM \citep[shaded area,][]{hatton2003} models. {\em Right}: point-like
sources, as compared to \citet{jarrett1994} predictions for spheroid 
(dashed line) and disk (dotted line) stars.} 
\label{fig:col_distr_mod}
\end{figure*}

The shaded regions in Fig. \ref{fig:counts_model} represent the locus of
predictions by the GalICS model, for the 10 different simulations 
considered. 
In the R band, consistency between model and observations holds at
the brighter fluxes, but the model seems to overpredict galaxy number counts 
by at least $\sim$30\% at magnitudes fainter than $\sim$20.
In the B band, the discrepancy is more significant, model predictions 
overpredicting number counts by a factor of $\sim 1.5$ at $B_J>23$. 

\citet{nagashima2002} presented a $\Lambda$CDM semi-analytic model, which produces
merger trees of Press-Schechter dark halos, in an $\Omega_m=0.3$,
$\Omega_\Lambda=0.7$, $h=0.7$ Universe.  This model includes 
the effects of dynamical response to supernova- and starburst-induced gas
removal on size and velocity dispersions, which seems to play an important role
in dwarf galaxy formation. 
The dotted line in Fig. \ref{fig:counts_model} represents the 
\citet{nagashima2002} model, which turns out to be fairly consistent with ESIS
data in all bands, excepted for some excess in the number counts at faint
fluxes.

Overall, the comparison of ESIS optical number counts to theoretical
predictions favors a scenario in which 
the Universe has experienced an epoch with
many encounters and interactions between galaxies, possibly at $z>1$, 
when the mean restframe UV-B luminosity of galaxies was larger than locally, 
because of enhanced star formation activity, likely triggered by mergers.

\subsection{Color distributions}

The distribution of observed colors, for both galaxies and stars,
provides additional constraints to cosmic evolution and 
galactic models.

The left panel in Fig. \ref{fig:col_distr_mod} shows the distribution of 
the $B_J-R_C$ color for extended sources, in different $B_J$ magnitude bins.
A non-evolutionary model \citep[long-dashed line,][]{roche1996} systematically
underpredicts the color counts, in the $22<B_J<24$ bins. The PLE model
\citep[short dashed,][]{roche1996} is more consistent with the data, but still
does not reproduce observations in the faintest magnitude bin. 
Concerning the two brightest magnitude bins, only few model predictions of the
observed-frame colors are available in literature.
The GalICS $\Lambda$-CDM color histogram seems to be too much peaked and to 
predict more sources than observed (see also Fig. \ref{fig:counts_model}), in
particular overestimating the number of faint blue galaxies.

The right panel in Fig. \ref{fig:col_distr_mod} compares the observed 
color distribution of point-like sources (defined as in Sect. 
\ref{sec:stellarity}) to the predictions of the \citet{jarrett1994} Milky Way
model. The dashed line refers to the spheroidal component, the dotted one to
the disk population. Despite model predictions are fairly consistent to the
observed point-like number counts (Fig. \ref{fig:counts}), the observed color 
distribution shows an excess of red sources at the faint fluxes and might 
caution against a residual contamination from unresolved galaxies.

\section{Potential for multi wavelength investigations}\label{sec:multiwave}

The ELAIS-S1 field benefits from an extensive follow up carried out over the whole
electromagnetic spectrum from X-rays to radio wavelengths (see also Fig.
\ref{fig:wfi_strategy}). In this Section we match ESIS data to catalogs in other
spectral domains, with the purpose of showing some of the potential of
multi-wavelength analysis of deep surveys.

\subsection{Observations in ELAIS-S1 and match to ESIS data}

The European Large-Area ISO Survey \citep[ELAIS,][]{rowanrobinson1999,oliver2000}, 
consisted of a deep, wide-angle survey at high galactic
latitudes, with the
Infrared Space Observatory (ISO). 
The primary ELAIS survey was carried out over 12 $[$deg$^2]$, at 15
$\mu$m \citep[with the ISOCAM camera,][]{cesarsky1996} and 90 $\mu$m 
\citep[with ISOPHOT,][]{lemke1996}; 
additional observations were performed in restricted areas at 6.7
and 175 $\mu$m, in collaboration with the FIRBACK team \citep{puget1999}. 
The main survey area was divided into three fields in the northern
hemisphere (N1, N2, and N3) and one field in the southern hemisphere (S1),
which is the target of the ESIS survey. 

The ELAIS-S1 field, centered at $\textrm{RA} = 00^{h}34^{m}44^{s}.4$,
$\textrm{DEC} = -43^\circ28'12''.0$ (J2000.0), covers a sky area of
about $2^\circ\times2^\circ$ and includes the minimum in Galactic 100 $\mu$m
cirrus emission in the Southern sky \citep{schlegel1998}.

The ELAIS survey in the S1 field reached 1.0, 0.7 and 70 $[$mJy$]$ depth in the
6.7, 15 and 90 $\mu$m bands \citep{rowanrobinson2004}. No 175 $\mu$m observation
was carried out in ELAIS-S1. The final 15 $\mu$m catalog contains 736 sources down
to $\sim$1 mJy, $\sim$20\% of which are stars \citep{vaccari2005}.

\citet{gruppioni1999} performed a radio (1.4 GHz) survey with the Australia
Telescope Compact Array (ATCA), 
over 4 sq. deg. in the ELAIS-S1 area, producing a catalog (already public)
of 652 sources, down to a minimum flux density of $\sim$0.2 mJy 
(5$\sigma$). 
The current 1.5 $[$deg$^2]$ ESIS area contains 268 of the ATCA radio sources 
detected in ELAIS-S1, $\sim$65\% of which have an optical counterpart.
A deeper survey is in progress by B.~Boyle and collaborators.

\citet{lafranca2004} presented spectroscopic and 
R-band data for 406 15 $\mu$m
sources in the ELAIS-S1 field, over the flux density range $0.5 < S(15\mu\textrm{m})
< 150$ $[$mJy$]$. The R band data were obtained by an ESO imaging campaign 
with the DFOSC instrument mounted on the 1.54 Danish/ESO telescope at La Silla
(Chile), reaching 95\% completeness level at R$\sim$22.5.

Spectroscopic observations of the optical counterparts of the ISOCAM S1 sources
were carried out at the 2dF/AAT, ESO Danish 1.5 m, 3.6 m and NTT telescopes,
adopting instrumental configurations with $\sim$10 \AA\ resolution and covering
the 4000-9000 \AA\ wavelength range on average \citep{lafranca2004}.

Spitzer/SWIRE observations in ELAIS-S1 cover roughly the whole ISO region and
a total sky area of $\sim$7 $[$deg$^2]$, reaching 5$\sigma$ sensitivities of 
3.7, 5.3, 48, 37.7 and 350 $\mu$Jy in the 3.6, 4.5, 5.8, 8.0, 24 $\mu$m channels
respectively \citep{lonsdale2004}. 
IRAC angular resolution spans 
$\sim$2.5 to 3 $[$arcsec$]$, while for the MIPS 24 $\mu$m band it is $\sim$6\arcsec.
Data reduction was carried out by the Spitzer Science Center and SWIRE team.
Absolute photometric uncertainty is $\sim$10\% both for all bands. SExtractor's 
Kron fluxes are considered for extended sources, while
aperture photometry is used for point-like objects, corrected for aperture
losses, as recommended by the Spitzer Science Center and SWIRE data release
documentation \citep{surace2004}. The astrometric accuracy of the SWIRE data
products is $\sim$0.5\arcsec (after image reconstruction and comparison to 2MASS
positions). The SWIRE band-merged catalog has been cross-correlated with 
ESIS WFI sources, by means of a simple nearest-match algorithm, adopting a
radius of 1\arcsec; 71779 SWIRE sources lie in the ESIS WFI current area; 
53267 matches (i.e. $\sim$74\%) are found using this criterion.

About 40\% of the ELAIS-S1 area was surveyed by BeppoSAX, down to a 2-10
keV sensitivity of $\sim$10$^{-13}$ $[$erg cm$^{-2}$s$^{-1}]$ 
\citep{alexander2001}.
More recent
XMM observations target the central area of the ISO field
(Puccetti et al., in preparation). A mosaic of four deep XMM pointings 
covers $\sim$0.6 $[$deg$^2]$ with the EPIC and two MOS cameras onboard 
XMM. The average net exposure time is $\sim$50 $[$ks$]$, in the energy range 
between 0.5 and 10 $[$keV$]$. 
WFI and XMM sources were cross-correlated if
lying within a radius of 2\arcsec (close to the typical 1$\sigma$ XMM error box); 
264 out of the 479 XMM sources are matched to ESIS objects.

The XMM ELAIS-S1 area is the target of further complementary observations. 
As part of ESO Large Programme 
170.A-0143 (P.I. A.~Cimatti, Dias et al., Buttery et al., in prep.), 
the ELAIS-S1 central $\sim$0.8 $[$deg$^2]$ was
the target of near-IR J and Ks imaging with SOFI on NTT, 
reaching J$<$21 and Ks$<$20 (Vega). 

VIMOS/VLT spectroscopy was carried out 
in summer 2004 and 2005, in $2\times 33$ hour observing runs
(73.A-0446 and 75.A-0428; PI: F. La
Franca). The whole XMM-NIR area was the target of low resolution 
spectroscopy over the wavelength range 5000-9500 \AA. 
During the first half of the survey, 1200 X-ray and $K_s<18.5$ galaxies were
observed; in period 75 the targets consisted of 1000 X-ray sources plus 24$\mu$m
SWIRE galaxies. R-band VLT pre-imaging provides additional data down to
$R\sim25$.
Further spectroscopy of optically bright ($R<21$) X-ray and mid-IR
sources is being performed at the 3.6m/ESO telescope, during Fall 2005.
These data will be presented in future papers by F.~La~Franca and collaborators.

\begin{figure*}[!t]
\centering
\includegraphics[width=0.85\textwidth]{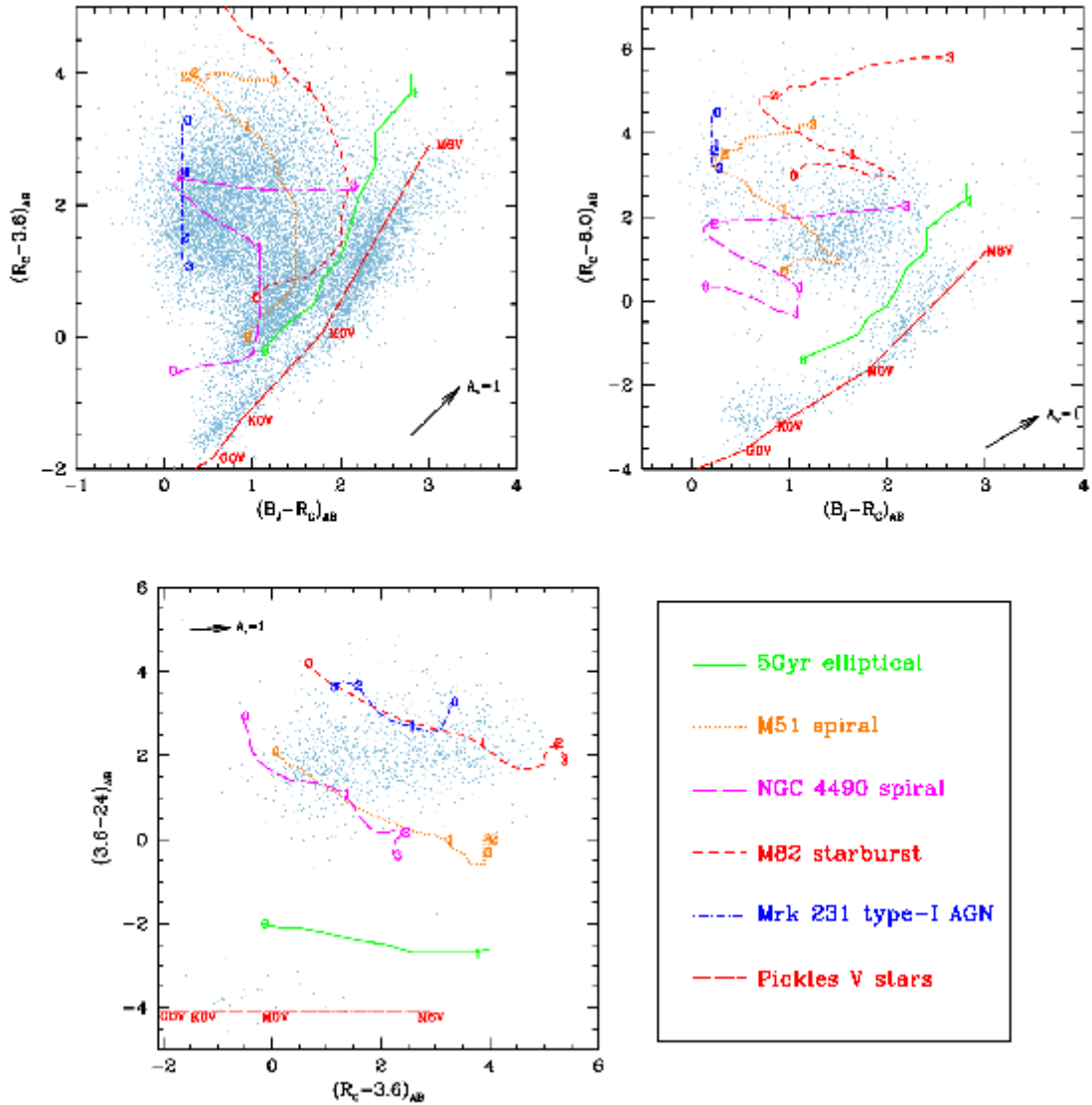}
\caption{Color-color plots for ESIS sources matched to SWIRE/Spitzer objects.
Template tracks represent k-corrected colors for different classes of
sources, as a function of redshift ($z=0-3$). The class-V stellar track 
was obtained by extending the \citet{pickles1998} library to the infrared,
assuming a Rayleigh-Jeans law. The reddening arrow refers to the
standard extinction law by \citet{cardelli1989}.}
\label{fig:opt_swire_col}
\end{figure*}

The ultraviolet Galaxy Evolution Explorer \citep[GALEX,][]{martin2005} Deep
Imaging Survey (DIS) is observing $\sim$80 $[$deg$^2]$ in 12 different areas,
including 
ELAIS-S1 \citep{burgarella2005}. Four ES1 GALEX tiles 
are already available to the public as part of data Release\footnote{GALEX Release 1 (GR1) DIS data were
released on Jan 4th, 2005 (http://www.galex.caltech.edu/)} no.1 (GR1); two of
these (namely tiles 00 and 01) match $\sim$0.93 $[$deg$^2]$ of the ESIS-WFI area
described here. The exposure times for these two tiles are $\sim$100 and
$\sim$50 $[$ks$]$ respectively.
The photometric catalogs in the far-UV (1344-1786 \AA) and near-UV (1771-2831
\AA) have been matched to ESIS data. In the ESIS-GR1 common area, 21025 GALEX sources
were detected. The in-flight measured point-source FWHM is 4\farcs0 and 5\farcs6 
in the two channels respectively. To be conservative, we used a matching radius
of 2\farcs0 only, which is sufficient for our demonstrative purposes of
showing the potential of observations in ELAIS-S1. 
In this way spurious and 
multiple matches are avoided. A more accurate match will be performed
in the future, in order to provide as complete a multi-$\lambda$ study as
possible. In this way, 17010 matches were found ($\sim$80\%). We use the flux
measurement provided by GR1.

\subsection{Optical-IR colors}

Color-color plots, built on a wide wavelength baseline turn out to be very useful 
to disentangle various classes of sources, different spectral domains
being sensitive to different emission components.

Figure \ref{fig:opt_swire_col} shows three color-color plots including optical WFI,
and mid-IR IRAC and MIPS photometric data. Synthetic tracks are overlaid on the
data, as obtained by k-correcting template spectral energy distributions (SEDs)
from $z=3$ to now.
The solid curve belongs to a 5 Gyr old elliptical model (truncated at $z=1.2$, to
be consistent with Universe age\footnote{$t_H\simeq$13.6 Gyr, for a $\Omega_m$=0.27,
$\Omega_\Lambda$=0.73, H$_0$=71 $[$km s$^{-1}$ Mpc$^{-1}]$ cosmology \citep{spergel2003}}), the
dotted line is a spiral M51-like model, the short-dashed line is a M82 prototype
starburst template; these three models are built with the GRASIL\footnote{GRASIL
homepage: http://web.pd.astro.it/granato/grasil/ grasil.html} code
\citep{silva1998}, 
and then upgraded by introducing observed PAH mid-IR features.
The long-dashed line belongs to a semi-empirical SED of the blue peculiar spiral
model of NGC 4490 (Polletta et al., in preparation) and the dot short-dash
model belongs to the type-I AGN Mrk 231 (Fritz et al., in prep.). Finally the
dot long-dash line belongs to class V stars 
and is obtained by extending the \citet{pickles1998} models to the infrared
assuming a Rayleigh-Jeans law. 

\begin{figure*}[!ht]
\centering
\includegraphics[width=0.85\textwidth]{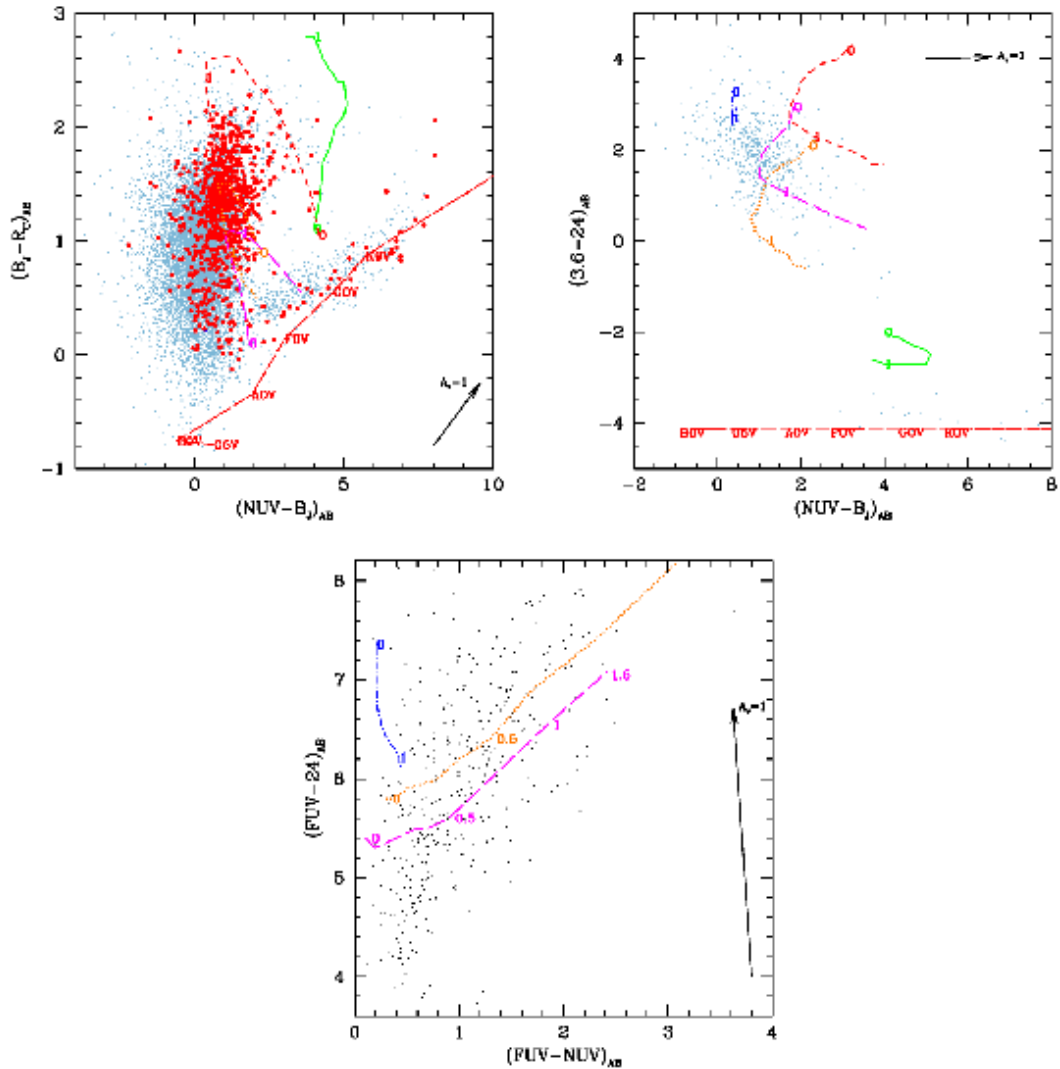}
\caption{Color-color plots for GALEX ultraviolet sources in the ESIS
current area. Template tracks are the same as in Fig. \ref{fig:opt_swire_col}. 
The thick dots in the left top panel represent sources detected in the 24 $\mu$m
Spitzer channel. The reddening arrow refers to the
standard extinction law by \citet{cardelli1989}.} 
\label{fig:galex_col}
\end{figure*}

IRAC data, when combined with the optical, turn out to be a  powerful tool 
to identify AGNs, normal galaxies, and stars. In the case of AGNs, the near-IR
emission is dominated by a dusty torus heated by the central engine, while for
normal galaxies, IRAC detects star light. Consequently, type-1 AGNs have flatter
optical-NIR slopes and bluer colors; these kind of sources occupy a small
characteristic locus in color space, while galaxy colors vary significantly as a
function of redshift. 
The optical-IRAC color-color diagrams are only partially effective in disentangling 
galaxies and stars. In fact, stars and ellipticals tend to have
similar colors at the $B-R$ reddest end, i.e. for high redshift ellipticals and
cool stars, because the spectral continuum in early-type galaxies is dominated by 
old stellar populations.

Those color-color diagrams that include MIPS 24 $\mu$m data would be potentially very
useful to further distinguish between different object classes, but SWIRE imaging is not
deep enough to detect many normal galaxies at 24 $\mu$m. This is particularly
true for elliptical galaxies.

\subsection{Ultraviolet sources}

Ultraviolet light is mainly emitted by young hot stars in galaxies, especially
during star formation events. Nonetheless, powerful starbursts are typically 
hosted by dusty, thick environments (molecular clouds); in these cases, the UV
emission of young stars is heavily extinguished and reprocessed to the mid- and
far-IR. 

The spectral energy distribution of type-1 AGNs contains
a significant feature in the far-ultraviolet to optical region, known as ``the
big blue bump''. This feature is usually explained as the thermal emission
from the optically thick accretion disk feeding the central massive black hole 
\citep[e.g.][]{shields1978,malkan1982}, which should peak in the 
extreme ultraviolet \citep[EUV, $100<\lambda<912$ \AA, e.g.][]{mathews1987}. 

The GALEX deep catalog in the ESIS area contains $\sim$19000 near-UV (1771-2831
\AA) and $\sim$9000 far-UV (1344-1786 \AA) sources. Roughly 7500 objects are
detected both in the NUV and FUV channels. 
GALEX flux densities are typically brighter than $\sim$1 $[\mu$Jy$]$ in both
bands. 

As a result of the cross-correlation between multiwavelength catalogs described
above, 856 sources turn out to be detected both by the NUV GALEX survey and the
MIPS 24 $\mu$m channel; 803 of these have a secure optical counterpart, detected
in all the three B, V, and R ESIS bands. 

The left top panel in Fig. \ref{fig:galex_col} represents the distribution of
sources in the $(B-R)$ vs. $(NUV-B)$ color space. Overlaid are the same template tracks
as in Fig. \ref{fig:opt_swire_col}, truncated at a redshift of 1.5.
Galaxies and stars are properly distinguished, the latter having $(NUV-B)>2$ and
$B-R<1.5$. The Type-1 AGN template occupy a very small area in color space, 
near to $(NUV-B)=(B-R)=0$. For redshifts smaller than 1, the starburst 
template (dashed line) traces  the area with $2<(NUV-B)<4$ and increasing
$(B-R)>1$;
at larger redshift, the $(B-R)$ color decreases again and the M82-like templates
intersects the locus of $z<1$ spiral galaxies.

Apart from stars, the bulk of observed sources lies blueward of normal spiral
galaxies. These are likely blue galaxies detected by GALEX thanks to some
moderately enhanced, unobscured star formation, or type-1 AGNs.
The locus of $z<1$ starbursts is not very populated, since dust extinction 
allows only the brightest sources to be detected in the UV.

The thicker points in the left top panel of Fig. \ref{fig:galex_col} represent
sources detected not only in the UV, but also at 24 $\mu$m by Spitzer.
The right top panel of Fig. \ref{fig:galex_col} compares the NUV and mid-IR excesses
for these objects and include also a few stars (at bottom).

It is worth noticing that, once more, the introduction of 24 $\mu$m data 
is critical in disentangling different kind of sources. 
Template tracks nicely trace 
the increase of ongoing star formation from normal spirals to starbursts. 

The bottom panel of Fig. \ref{fig:galex_col} shows the ratio of mid-IR and
far-UV fluxes as a function of the UV color of GALEX-24$\mu$m sources. 
The M82 starburst template lies above $(FUV-24)>8$:  
GALEX is not sensitive to young stars enshrouded in thick 
dusty clouds, which instead are detected in the mid-IR domain, thanks to dust
reprocessing of the extinguished light. 
Blue UV galaxies with some 24 $\mu$m excess emission 
populate the upper part of the diagram (reddest $[FUV-24]$), in between the
normal spirals and starbursts loci. 

These color-color diagrams suggest that the majority NUV+MIPS sources are low or
moderate redshift late-type galaxies ($z\le0.5$), with enhanced star formation
activity. This ongoing
star formation is almost unextinguished, or moderately absorbed ($A_V<0.5$), 
since both an ultraviolet and a mid-infrared excesses are detected in the SEDs
of these galaxies.


\subsection{SEDs examples}

\begin{figure*}[!t]
\centering
\includegraphics[width=0.8\textwidth]{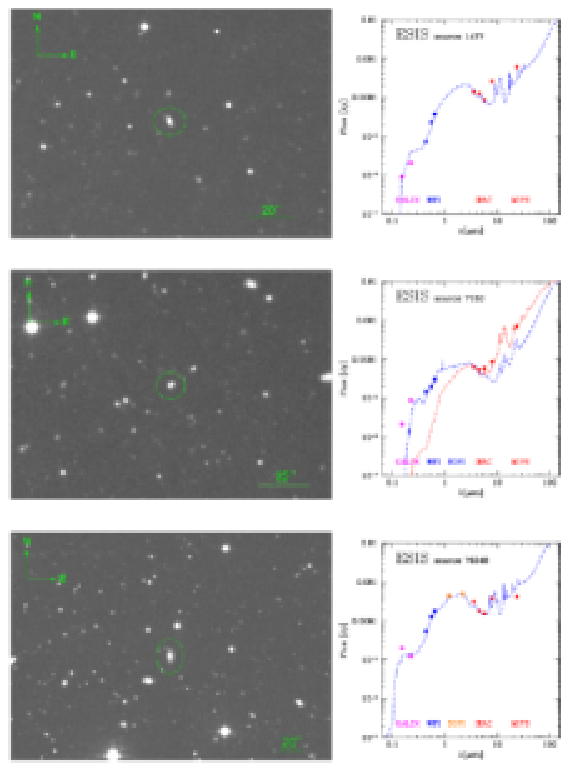}
\caption{Spectral energy distributions of three sample sources, from the UV to
the mid-IR. The superimposed templates are: 1477, spiral galaxy at $z=0.5$; 7160, 
starburst (dot-dashed line) and blue spiral (dashed) at $z=0.8$, 76348 spiral galaxy at $z=0.18$.} 
\label{fig:sed1}
\end{figure*}

\begin{figure*}[!ht]
\centering
\includegraphics[width=0.8\textwidth]{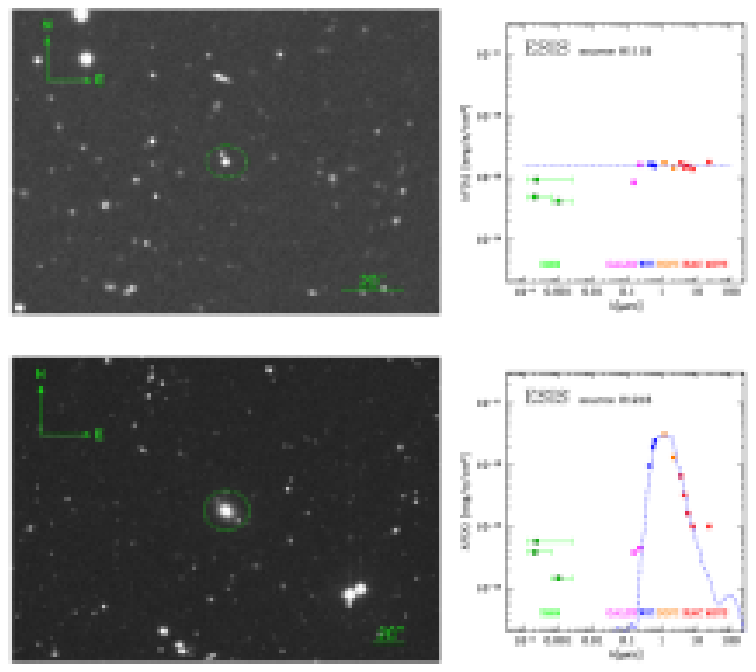}
\caption{Spectral energy distributions of two X-ray detected sources.
Object 81115 has a typical type-1 AGN SED; the optical emission from source 81218 
is reproduced by an elliptical galaxy template at $z=0.2$ (dashed line).} 
\label{fig:sed3}
\end{figure*}

Multiwavelength coverage from the X-rays to the far-IR allows the study of the
nature of individual sources. 

Blue optical restframe emission is dominated by young stars, while redwards (V, R
bands), moderate-aged (type A-F) stars are preferentially detected. 

Near-IR
imaging traces the old stellar population of normal galaxies, and at
$\lambda>4\ \mu$m, hot ($T\sim1000$ $[$K$]$) dust emission, if present,
is no longer negligible (e.g. in a AGN-heated torus). 

Mid-IR radiation (5-30 $\mu$m) is dominated by warm ($T\sim100-300$ $[$K$]$)
dust, either heated by young stars in starburst galaxies or
belonging to AGN tori. Polycyclic Aromatic Hydrocarbons molecules
characterize mid-IR restframe light: bright emission bands are detected at 
3.3, 6.2, 7.7, 8.6 and 11.3 $\mu$m and are typical of starburst galaxies.

In the far-IR spectral domain (30-200 $\mu$m), 
warm ($T\sim100-300$ $[$K$]$) and cold ($T\sim40$ $[$K$]$, belonging to cirrus)
dust contribute to the energy budget, producing very bright luminosities
in active galaxies (starbursts, AGNs) as well as in normal spirals. 

Ultraviolet emission comes from young stars, but is extinguished 
by dust and reprocessed to the infrared (see also previous Section).
Elliptical galaxies produce a significant ``UV excess'', with regards to old stellar
continuum, due to the presence of evolved components such as planetary
nebulae and AGB manqu\`e stars \citep[see, for example,][]{oconnell1999}.

In the X-rays, AGN power-law emission is detected, either
direct (type-I sources) or partially absorbed (type-II) in addition to scattered light
(if the optical depth is high).
Starburst galaxies host several X-ray components \citep[see, e.g.,][]{persic2002}, 
most prominently thermal emission by X-ray binaries.
Hot plasma in galaxy clusters produces extended thermal bremsstrahlung X-ray
radiation. 

Finally, adding deep radio observations will provide further insights into the nature of
the ESIS galaxy populations, detecting synchrotron radiation from starbursts and
radio loud AGN.

Figures \ref{fig:sed1} and \ref{fig:sed3} show the SEDs of five ESIS nearby
sources, exemplifying different classes of objects. 
None of these sources were detected by ISO, nor by SWIRE's 70 and 160 $\mu$m
observations.

Source no. {\bf 1477} is a typical spiral galaxy with some enhanced,
absorbed, ongoing star formation, displaying a moderate mid-IR excess in Spitzer's
8.0 and 24 $\mu$m channels, and red UV-optical colors.
Overlaid is a M51 spiral template shifted to $z=0.5$.

Source no. {\bf 7160}'s SED resembles a blue starburst, showing a bright IR excess,
likely powered by warm dust and PAH features, and moderately blue optical colors.
The overlaid dashed line is a blue spiral template based on the nearby galaxy
NGC4490 (Polletta et al., in prep.) shifted to $z=0.8$, while the 
dot-dashed line represents a M82 template at the same redshift, which 
reproduces the observed IR excess fairly well. 
The M82 template emission seems to be 
too absorbed in the optical-UV, being systematically fainter than 
the observed fluxes. This effect indicates that this source probably hosts 
a moderately extinguished starburst.

Object no. {\bf 76348} is a normal spiral galaxy, with some weak
unextinguished ongoing star formation producing some UV excess. This source lies
in the area observed with SOFI in the J and Ks bands. The dashed line 
is a M100 template, redshifted to $z=0.18$.

Figure \ref{fig:sed3} shows two objects detected in the X-rays and 
in the near-IR: a type-I AGN
({\bf 81115}) and a spheroidal galaxy ({\bf 81218}).
The former SED is a flat power-law spectrum ($\lambda F(\lambda)=\textrm{{\em
const}}$) over the entire wavelength range 
considered.
The latter source shows some 24 $\mu$m emission in excess to what is predicted
for a simple old stellar population (the dashed line represents an elliptical 
galaxy template at $z=0.2$), possibly powered by a hidden AGN, which is
responsible also for the X-ray flux. 
Note that the presumably high UV-optical extinction 
of the possible type-2 AGN is confirmed by the quite
hard X-ray spectrum.
These plots report $\lambda F(\lambda)$ instead of flux
density, in order to better highlight the energy budget in the X-rays, compared
to the other wavelengths.

\section{Summary}

Optical B, V, R imaging of 1.5 square degree
in the ELAIS-S1 area, belonging to the ESIS survey has been presented. 
The data consist of deep Wide Field Imager (2.2m ESO/MPI telescope in La Silla)
observations, requiring $\ge$30 different frames per WFI pointing, for a total
exposure time of 2.5h per band. Ad hoc technical solutions have been developed
for correcting illumination gradients by means of {\em super-sky flat} fields,
and to remove color effects and astrometric distortions.
Accurate simulations have been run in order to estimate source extraction
efficiency and photometric uncertainties.

A total of 132712 sources are detected on the 1.5 $[$deg$^2]$ area analyzed
here. Equatorial coordinates of objects are constrained within a $\sim$0.15
$[$arcsec$]$ r.m.s. uncertainty, with respect to the GSC 2.2 catalog;
flux errors are estimated to be of the order of 
$\sim$2, 10, 20\% at mag. 20, 23, 24; we reach 95\% completeness
B,V$\sim$25 and R$\sim$24.5.

Number counts in the B and R bands were compared to literature data and
theoretical models of galaxy evolution. Our data are consistent with
previous works from various surveys \citep[e.g. HDFN, HDFS,
VIRMOS,][]{williams1996,metcalfe2001,mccracken2003}.
The stellar contribution to the counts has been taken into account 
both using SExtractor's stellarity index and the 
\citet{jarrett1994} Milky Way model.
Comparison to number count models by \citet{metcalfe2001}, 
semi-analytical $\Lambda$CDM mock catalogs \citep[GalICS][]{hatton2003}
and $\Lambda$CDM \citep{nagashima2002} predictions
confirm the need of galaxy luminosity evolution to properly reproduce the
observed data. 
ESIS galaxy number counts suggest that at $z>1$ 
the rate of galaxy encounters was probably larger than today, and the
mean restframe UV-Blue luminosity of galaxies was larger than locally, because
of enhanced star formation activity, possibly triggered by mergers. 

The {\em ESO-Spitzer Imaging extragalactic Survey}
is providing optical identifications, colors, rough
morphologies, photometric redshifts, for a good fraction of the  $\sim 300000$
IR sources detected 
by SWIRE/Spitzer ($\lambda=$3.6, 4.5, 5.8. 8.0, 24, 70, 160 $\mu$m) in the 5
$[$deg$^2]$ ELAIS-S1 area.  

In addition to the Spitzer/SWIRE survey, ancillary data include 
observations in the X-rays (XMM, Puccetti et al., in prep.), ultra-violet 
\citep[GALEX,][]{martin2005}, near-IR (NTT, Dias et al., Buttery et al., in prep.), 
and radio \citep[ATCA,][]{gruppioni1999} spectral domains.
Some of the enormous potential stored in this multi-wavelength dataset has been
traced in the last Section of this paper. 

Thanks to the wide wavelength baseline covered, Spitzer-optical color-color diagrams 
turn out to be very powerful tools to disentangle different source populations,
such as normal spirals, elliptical galaxies, starbursts and active galactic nuclei.

The ultraviolet to mid-IR color space was analyzed, matching ESIS sources to the 
GALEX Deep Imaging Survey catalog released in January 2005. Roughly 80\% of the UV sources
in the ESIS area are detected in the optical, while only a small fraction have a mid-IR
24 $\mu$m Spitzer counterpart. 
UV-optical and UV-24$ \mu$m color-color diagrams suggest that sources 
detected both by GALEX in the ultraviolet and by MIPS at 24 $\mu$m are low or
moderate redshift late-type galaxies, with enhanced, 
almost unextinguished ongoing star formation activity.

Finally, X-ray to far-IR spectral energy distributions of some ESIS sources 
were presented as examples of the characterization 
of SWIRE objects, based on ancillary data.

Such a comprehensive wavelength coverage of a large area will give a significant
contribution, among others, in: {\em (1)} 
deriving the physical properties (e.g. ongoing rate of star formation and
assembled stellar mass) of Spitzer galaxies up to $z\sim3$, {\em (2)} searching for
distant ($z>1$) galaxy clusters, 
{\em (3)} studying the statistical properties (e.g. number counts) of starburst, evolved
galaxies, AGNs and comparing them to theoretical models, {\em (4)} studying the cosmic
star formation density as a function of redshift, {\em (5)} building the luminosity and
mass functions of galaxies and studying their dependence on redshift and
environment. 

Updates on the ESIS project are available on the web page {\tt
http://dipastro.pd.astro.it/esis}. The data described in this paper 
are included in the third SWIRE Spitzer Legacy data release (Fall 2005).

\begin{acknowledgements}

We wish to thank the referee, S. Arnouts, for his very constructive suggestions.
We would like to thank the whole SWIRE team for preparing the ES1 Spitzer data
and Ezio Pignatelli for useful discussions about SExtractor. 
SB was supported by Uni-PD and INAF-OaPD grants, ASI contract no. I/R/062/02, 
and ESO DGDF 2004 studentships.

The Spitzer Space Telescope
is operated by the Jet Propulsion Laboratory, California Institute of Technology, under
contract with NASA. 
SWIRE is supported by NASA through the SIRTF Legacy
Program under contract 1407 with the Jet Propulsion Laboratory.

We acknowledge NASA's support for construction,
operation, and science analysis for the GALEX mission, developed in cooperation
with the Centre National d'Etudes Spatiales of France and the Korean Ministry of
Science and Technology. 

XMM-Newton is an ESA science mission with instruments and contributions
directly funded by ESA Member States and the USA (NASA). 

This work uses the GalICS/MoMaF Database of Galaxies (http://galics.iap.fr).

\end{acknowledgements}


\bibliographystyle{aa}
\bibliography{4548bib}   

\begin{thebibliography}{75}
\expandafter\ifx\csname natexlab\endcsname\relax\def\natexlab#1{#1}\fi

\bibitem[{{Alcal{\' a}} {et~al.}(2004){Alcal{\' a}}, {Pannella}, {Puddu},
  {Radovich}, {Silvotti}, {Arnaboldi}, {Capaccioli}, {Covone}, {Dall'Ora}, {De
  Lucia}, {Grado}, {Longo}, {Mercurio}, {Musella}, {Napolitano}, {Pavlov},
  {Rifatto}, {Ripepi}, \& {Scaramella}}]{alcala2004}
{Alcal{\' a}}, J.~M., {Pannella}, M., {Puddu}, E., {et~al.} 2004, \aap, 428,
  339

\bibitem[{{Alexander} {et~al.}(2001){Alexander}, {La Franca}, {Fiore},
  {Barcons}, {Ciliegi}, {Danese}, {Della Ceca}, {Franceschini}, {Gruppioni},
  {Matt}, {Matute}, {Oliver}, {Pompilio}, {Wolter}, {Efstathiou},
  {H{\'e}raudeau}, {Perola}, {Perri}, {Rigopoulou}, {Rowan-Robinson}, \&
  {Serjeant}}]{alexander2001}
{Alexander}, D.~M., {La Franca}, F., {Fiore}, F., {et~al.} 2001, \apj, 554, 18

\bibitem[{{Arnouts} {et~al.}(1999){Arnouts}, {D'Odorico}, {Cristiani},
  {Zaggia}, {Fontana}, \& {Giallongo}}]{arnouts1999}
{Arnouts}, S., {D'Odorico}, S., {Cristiani}, S., {et~al.} 1999, \aap, 341, 641

\bibitem[{{Arnouts} {et~al.}(2001){Arnouts}, {Vandame}, {Benoist},
  {Groenewegen}, {da Costa}, {Schirmer}, {Mignani}, {Slijkhuis},
  {Hatziminaoglou}, {Hook}, {Madejsky}, {Rit{\'e}}, \& {Wicenec}}]{arnouts2001}
{Arnouts}, S., {Vandame}, B., {Benoist}, C., {et~al.} 2001, \aap, 379, 740

\bibitem[{{Baade} {et~al.}(1999){Baade}, {Meisenheimer}, {Iwert}, {Alonso},
  {Augusteijn}, {Beletic}, {Bellemann}, {Benesch}, {Boehm}, {Boehnhardt},
  {Brewer}, {Deiries}, {Delabre}, {Donaldson}, {Dupuy}, {Franke}, {Gerdes},
  {Gilliotte}, {Grimm}, {Haddad}, {Hess}, {Ihle}, {Klein}, {Lenzen}, {Lizon},
  {Mancini}, {Muench}, {Pizarro}, {Prado}, {Rahmer}, {Reyes}, {Richardson},
  {Robledo}, {Sanchez}, {Silber}, {Sinclaire}, {Wackermann}, \&
  {Zaggia}}]{baade1999}
{Baade}, D., {Meisenheimer}, K., {Iwert}, O., {et~al.} 1999, The Messenger, 95,
  15

\bibitem[{{Bertin} \& {Arnouts}(1996)}]{bertin1996}
{Bertin}, E. \& {Arnouts}, S. 1996, \aaps, 117, 393

\bibitem[{{Blaizot} {et~al.}(2005){Blaizot}, {Wadadekar}, {Guiderdoni},
  {Colombi}, {Bertin}, {Bouchet}, {Devriendt}, \& {Hatton}}]{blaizot2005}
{Blaizot}, J., {Wadadekar}, Y., {Guiderdoni}, B., {et~al.} 2005, \mnras, 360,
  159

\bibitem[{{Blumenthal} {et~al.}(1984){Blumenthal}, {Faber}, {Primack}, \&
  {Rees}}]{blumenthal1984}
{Blumenthal}, G.~R., {Faber}, S.~M., {Primack}, J.~R., \& {Rees}, M.~J. 1984,
  \nat, 311, 517

\bibitem[{{Bruzual} \& {Charlot}(1993)}]{bruzual1993}
{Bruzual}, G. \& {Charlot}, S. 1993, \apj, 405, 538

\bibitem[{{Burgarella} {et~al.}(2005){Burgarella}, {Buat}, {Small}, {Barlow},
  {Boissier}, {Gil de Paz}, {Heckman}, {Madore}, {Martin}, {Rich}, {Bianchi},
  {Byun}, {Donas}, {Forster}, {Friedman}, {Jelinsky}, {Lee}, {Malina},
  {Milliard}, {Morrissey}, {Neff}, {Schiminovich}, {Siegmund}, {Szalay},
  {Welsh}, \& {Wyder}}]{burgarella2005}
{Burgarella}, D., {Buat}, V., {Small}, T., {et~al.} 2005, \apjl, 619, L63

\bibitem[{{Calabretta} \& {Greisen}(2002)}]{calabretta2002}
{Calabretta}, M.~R. \& {Greisen}, E.~W. 2002, \aap, 395, 1077

\bibitem[{{Cardelli} {et~al.}(1989){Cardelli}, {Clayton}, \&
  {Mathis}}]{cardelli1989}
{Cardelli}, J.~A., {Clayton}, G.~C., \& {Mathis}, J.~S. 1989, \apj, 345, 245

\bibitem[{{Cesarsky} {et~al.}(1996){Cesarsky}, {Abergel}, {Agnese}, {Altieri},
  {Augueres}, {Aussel}, {Biviano}, {Blommaert}, {Bonnal}, {Bortoletto},
  {Boulade}, {Boulanger}, {Cazes}, {Cesarsky}, {Chedin}, {Claret}, {Combes},
  {Cretolle}, {Davies}, {Desert}, {Elbaz}, {Engelmann}, {Epstein},
  {Franceschini}, {Gallais}, {Gastaud}, {Gorisse}, {Guest}, {Hawarden},
  {Imbault}, {Kleczewski}, {Lacombe}, {Landriu}, {Lapegue}, {Lena}, {Longair},
  {Mandolesi}, {Metcalfe}, {Mosquet}, {Nordh}, {Okumura}, {Ott}, {Perault},
  {Perrier}, {Persi}, {Puget}, {Purkins}, {Rio}, {Robert}, {Rouan}, {Roy},
  {Saint-Pe}, {Sam Lone}, {Sargent}, {Sauvage}, {Sibille}, {Siebenmorgen},
  {Sirou}, {Soufflot}, {Starck}, {Tiphene}, {Tran}, {Ventura}, {Vigroux},
  {Vivares}, \& {Wade}}]{cesarsky1996}
{Cesarsky}, C.~J., {Abergel}, A., {Agnese}, P., {et~al.} 1996, \aap, 315, L32

\bibitem[{{Cimatti}(2003)}]{cimatti2003}
{Cimatti}, A. 2003, in The Mass of Galaxies at Low and High Redshift, 124

\bibitem[{{Daddi} {et~al.}(2004){Daddi}, {Cimatti}, {Renzini}, {Vernet},
  {Conselice}, {Pozzetti}, {Mignoli}, {Tozzi}, {Broadhurst}, {di Serego
  Alighieri}, {Fontana}, {Nonino}, {Rosati}, \& {Zamorani}}]{daddi2004}
{Daddi}, E., {Cimatti}, A., {Renzini}, A., {et~al.} 2004, \apjl, 600, L127

\bibitem[{{Dickinson} {et~al.}(2003){Dickinson}, {Papovich}, {Ferguson}, \&
  {Budav{\' a}ri}}]{dickinson2003}
{Dickinson}, M., {Papovich}, C., {Ferguson}, H.~C., \& {Budav{\' a}ri}, T.
  2003, \apj, 587, 25

\bibitem[{{D'Odorico} {et~al.}(2003){D'Odorico}, {Aguayo}, {Brillant},
  {Canavan}, {Castillo}, {Cuby}, {Dekker}, {Haddad}, {Izzo}, {Hanuschik},
  {Kissler-Patig}, {Lizon}, {Marchesi}, {Marconi}, {Moller}, {Palsa}, {Robert},
  {Romaniello}, \& {Sartoretti}}]{dodorico2003}
{D'Odorico}, S., {Aguayo}, A.-M., {Brillant}, S., {et~al.} 2003, The Messenger,
  113, 26

\bibitem[{{Drory} {et~al.}(2005){Drory}, {Salvato}, {Gabasch}, {Bender},
  {Hopp}, {Feulner}, \& {Pannella}}]{drory2005}
{Drory}, N., {Salvato}, M., {Gabasch}, A., {et~al.} 2005, \apjl, 619, L131

\bibitem[{{Eggen} {et~al.}(1962){Eggen}, {Lynden-Bell}, \&
  {Sandage}}]{eggen1962}
{Eggen}, O.~J., {Lynden-Bell}, D., \& {Sandage}, A.~R. 1962, \apj, 136, 748

\bibitem[{{Elbaz} {et~al.}(2002){Elbaz}, {Cesarsky}, {Chanial}, {Aussel},
  {Franceschini}, {Fadda}, \& {Chary}}]{elbaz2002}
{Elbaz}, D., {Cesarsky}, C.~J., {Chanial}, P., {et~al.} 2002, \aap, 384, 848

\bibitem[{{Ellis}(1997)}]{ellis1997}
{Ellis}, R.~S. 1997, \araa, 35, 389

\bibitem[{{Fazio} {et~al.}(2004){Fazio}, {Hora}, {Allen}, {Ashby}, {Barmby},
  {Deutsch}, {Huang}, {Kleiner}, {Marengo}, {Megeath}, {Melnick}, {Pahre},
  {Patten}, {Polizotti}, {Smith}, {Taylor}, {Wang}, {Willner}, {Hoffmann},
  {Pipher}, {Forrest}, {McMurty}, {McCreight}, {McKelvey}, {McMurray}, {Koch},
  {Moseley}, {Arendt}, {Mentzell}, {Marx}, {Losch}, {Mayman}, {Eichhorn},
  {Krebs}, {Jhabvala}, {Gezari}, {Fixsen}, {Flores}, {Shakoorzadeh}, {Jungo},
  {Hakun}, {Workman}, {Karpati}, {Kichak}, {Whitley}, {Mann}, {Tollestrup},
  {Eisenhardt}, {Stern}, {Gorjian}, {Bhattacharya}, {Carey}, {Nelson},
  {Glaccum}, {Lacy}, {Lowrance}, {Laine}, {Reach}, {Stauffer}, {Surace},
  {Wilson}, {Wright}, {Hoffman}, {Domingo}, \& {Cohen}}]{fazio2004}
{Fazio}, G.~G., {Hora}, J.~L., {Allen}, L.~E., {et~al.} 2004, \apjs, 154, 10

\bibitem[{{Fontana} {et~al.}(2004){Fontana}, {Pozzetti}, {Donnarumma},
  {Renzini}, {Cimatti}, {Zamorani}, {Menci}, {Daddi}, {Giallongo}, {Mignoli},
  {Perna}, {Salimbeni}, {Saracco}, {Broadhurst}, {Cristiani}, {D'Odorico}, \&
  {Gilmozzi}}]{fontana2004}
{Fontana}, A., {Pozzetti}, L., {Donnarumma}, I., {et~al.} 2004, \aap, 424, 23

\bibitem[{{Franceschini} {et~al.}(2001){Franceschini}, {Aussel}, {Cesarsky},
  {Elbaz}, \& {Fadda}}]{franceschini2001}
{Franceschini}, A., {Aussel}, H., {Cesarsky}, C.~J., {Elbaz}, D., \& {Fadda},
  D. 2001, \aap, 378, 1

\bibitem[{{Franceschini} {et~al.}(2003){Franceschini}, {Berta}, {Rigopoulou},
  {Aussel}, {Cesarsky}, {Elbaz}, {Genzel}, {Moy}, {Oliver}, {Rowan-Robinson},
  \& {Van der Werf}}]{franceschini2003}
{Franceschini}, A., {Berta}, S., {Rigopoulou}, D., {et~al.} 2003, \aap, 403,
  501

\bibitem[{{Gardner} {et~al.}(1996){Gardner}, {Sharples}, {Carrasco}, \&
  {Frenk}}]{gardner1996}
{Gardner}, J.~P., {Sharples}, R.~M., {Carrasco}, B.~E., \& {Frenk}, C.~S. 1996,
  \mnras, 282, L1

\bibitem[{{Gruppioni} {et~al.}(1999){Gruppioni}, {Ciliegi}, {Rowan-Robinson},
  {Cram}, {Hopkins}, {Cesarsky}, {Danese}, {Franceschini}, {Genzel},
  {Lawrence}, {Lemke}, {McMahon}, {Miley}, {Oliver}, {Puget}, \&
  {Rocca-Volmerange}}]{gruppioni1999}
{Gruppioni}, C., {Ciliegi}, P., {Rowan-Robinson}, M., {et~al.} 1999, \mnras,
  305, 297

\bibitem[{{Guiderdoni} {et~al.}(1998){Guiderdoni}, {Hivon}, {Bouchet}, \&
  {Maffei}}]{guiderdoni1998}
{Guiderdoni}, B., {Hivon}, E., {Bouchet}, F.~R., \& {Maffei}, B. 1998, \mnras,
  295, 877

\bibitem[{{Guiderdoni} \& {Rocca-Volmerange}(1991)}]{guiderdoni1991}
{Guiderdoni}, B. \& {Rocca-Volmerange}, B. 1991, \aap, 252, 435

\bibitem[{{Hatton} {et~al.}(2003){Hatton}, {Devriendt}, {Ninin}, {Bouchet},
  {Guiderdoni}, \& {Vibert}}]{hatton2003}
{Hatton}, S., {Devriendt}, J.~E.~G., {Ninin}, S., {et~al.} 2003, \mnras, 343,
  75

\bibitem[{{Hauser} {et~al.}(1998){Hauser}, {Arendt}, {Kelsall}, {Dwek},
  {Odegard}, {Weiland}, {Freudenreich}, {Reach}, {Silverberg}, {Moseley},
  {Pei}, {Lubin}, {Mather}, {Shafer}, {Smoot}, {Weiss}, {Wilkinson}, \&
  {Wright}}]{hauser1998}
{Hauser}, M.~G., {Arendt}, R.~G., {Kelsall}, T., {et~al.} 1998, \apj, 508, 25

\bibitem[{{Jarrett}(1992)}]{jarrett1992}
{Jarrett}, T.~H. 1992, Ph.D.~Thesis

\bibitem[{{Jarrett} {et~al.}(1994){Jarrett}, {Dickman}, \&
  {Herbst}}]{jarrett1994}
{Jarrett}, T.~H., {Dickman}, R.~L., \& {Herbst}, W. 1994, \apj, 424, 852

\bibitem[{{Kauffmann} \& {Charlot}(1998)}]{kauffmann1998}
{Kauffmann}, G. \& {Charlot}, S. 1998, \mnras, 297, L23

\bibitem[{{La Franca} {et~al.}(2004){La Franca}, {Gruppioni}, {Matute},
  {Pozzi}, {Lari}, {Mignoli}, {Zamorani}, {Alexander}, {Cocchia}, {Danese},
  {Franceschini}, {H{\'e}raudeau}, {Kotilainen}, {Linden-V{\o}rnle}, {Oliver},
  {Rowan-Robinson}, {Serjeant}, {Spinoglio}, \& {Verma}}]{lafranca2004}
{La Franca}, F., {Gruppioni}, C., {Matute}, I., {et~al.} 2004, \aj, 127, 3075

\bibitem[{{Landolt}(1992)}]{landolt1992}
{Landolt}, A.~U. 1992, \aj, 104, 340

\bibitem[{{Le F{\` e}vre} {et~al.}(2002){Le F{\` e}vre}, {Mancini}, {Saisse},
  {Brau-Nogu{\' e}}, {Caputi}, {Castinel}, {D'Odorico}, {Garilli}, {Kissler},
  {Lucuix}, {Mancini}, {Pauget}, {Sciarretta}, {Scodeggio}, {Tresse},
  {Maccagni}, {Picat}, \& {Vettolani}}]{lefevre2002}
{Le F{\` e}vre}, O., {Mancini}, D., {Saisse}, M., {et~al.} 2002, The Messenger,
  109, 21

\bibitem[{{Lemke} {et~al.}(1996){Lemke}, {Klaas}, {Abolins}, {Abraham},
  {Acosta-Pulido}, {Bogun}, {Castaneda}, {Cornwall}, {Drury}, {Gabriel},
  {Garzon}, {Gemuend}, {Groezinger}, {Gruen}, {Haas}, {Hajduk}, {Hall},
  {Heinrichsen}, {Herbstmeier}, {Hirth}, {Joseph}, {Kinkel}, {Kirches},
  {Koempe}, {Kraetschmer}, {Kreysa}, {Krueger}, {Kunkel}, {Laureijs},
  {Luetzow-Wentzky}, {Mattila}, {Mueller}, {Pacher}, {Pelz}, {Popow},
  {Rasmussen}, {Rodriguez Espinosa}, {Richards}, {Russell}, {Schnopper},
  {Schubert}, {Schulz}, {Telesco}, {Tilgner}, {Tuffs}, {Voelk}, {Walker},
  {Wells}, \& {Wolf}}]{lemke1996}
{Lemke}, D., {Klaas}, U., {Abolins}, J., {et~al.} 1996, \aap, 315, L64

\bibitem[{{Lonsdale} {et~al.}(2004){Lonsdale}, {Polletta}, {Surace}, {Shupe},
  {Fang}, {Xu}, {Smith}, {Siana}, {Rowan-Robinson}, {Babbedge}, {Oliver},
  {Pozzi}, {Davoodi}, {Owen}, {Padgett}, {Frayer}, {Jarrett}, {Masci},
  {O'Linger}, {Conrow}, {Farrah}, {Morrison}, {Gautier}, {Franceschini},
  {Berta}, {Perez-Fournon}, {Hatziminaoglou}, {Afonso-Luis}, {Dole}, {Stacey},
  {Serjeant}, {Pierre}, {Griffin}, \& {Puetter}}]{lonsdale2004}
{Lonsdale}, C., {Polletta}, M.~d.~C., {Surace}, J., {et~al.} 2004, \apjs, 154,
  54

\bibitem[{{Lonsdale} {et~al.}(2003){Lonsdale}, {Smith}, {Rowan-Robinson},
  {Surace}, {Shupe}, {Xu}, {Oliver}, {Padgett}, {Fang}, {Conrow},
  {Franceschini}, {Gautier}, {Griffin}, {Hacking}, {Masci}, {Morrison},
  {O'Linger}, {Owen}, {P{\' e}rez-Fournon}, {Pierre}, {Puetter}, {Stacey},
  {Castro}, {Del Carmen Polletta}, {Farrah}, {Jarrett}, {Frayer}, {Siana},
  {Babbedge}, {Dye}, {Fox}, {Gonzalez-Solares}, {Salaman}, {Berta}, {Condon},
  {Dole}, \& {Serjeant}}]{lonsdale2003}
{Lonsdale}, C.~J., {Smith}, H.~E., {Rowan-Robinson}, M., {et~al.} 2003, \pasp,
  115, 897

\bibitem[{{Malkan} \& {Sargent}(1982)}]{malkan1982}
{Malkan}, M.~A. \& {Sargent}, W.~L.~W. 1982, \apj, 254, 22

\bibitem[{{Martin} {et~al.}(2005){Martin}, {Fanson}, {Schiminovich},
  {Morrissey}, {Friedman}, {Barlow}, {Conrow}, {Grange}, {Jelinsky},
  {Milliard}, {Siegmund}, {Bianchi}, {Byun}, {Donas}, {Forster}, {Heckman},
  {Lee}, {Madore}, {Malina}, {Neff}, {Rich}, {Small}, {Surber}, {Szalay},
  {Welsh}, \& {Wyder}}]{martin2005}
{Martin}, D.~C., {Fanson}, J., {Schiminovich}, D., {et~al.} 2005, \apjl, 619,
  L1

\bibitem[{{Mathews} \& {Ferland}(1987)}]{mathews1987}
{Mathews}, W.~G. \& {Ferland}, G.~J. 1987, \apj, 323, 456

\bibitem[{{McCracken} {et~al.}(2003){McCracken}, {Radovich}, {Bertin},
  {Mellier}, {Dantel-Fort}, {Le F{\` e}vre}, {Cuillandre}, {Gwyn}, {Foucaud},
  \& {Zamorani}}]{mccracken2003}
{McCracken}, H.~J., {Radovich}, M., {Bertin}, E., {et~al.} 2003, \aap, 410, 17

\bibitem[{{Metcalfe} {et~al.}(2001){Metcalfe}, {Shanks}, {Campos}, {McCracken},
  \& {Fong}}]{metcalfe2001}
{Metcalfe}, N., {Shanks}, T., {Campos}, A., {McCracken}, H.~J., \& {Fong}, R.
  2001, \mnras, 323, 795

\bibitem[{{Metcalfe} {et~al.}(1991){Metcalfe}, {Shanks}, {Fong}, \&
  {Jones}}]{metcalfe1991}
{Metcalfe}, N., {Shanks}, T., {Fong}, R., \& {Jones}, L.~R. 1991, \mnras, 249,
  498

\bibitem[{{Metcalfe} {et~al.}(1995){Metcalfe}, {Shanks}, {Fong}, \&
  {Roche}}]{metcalfe1995}
{Metcalfe}, N., {Shanks}, T., {Fong}, R., \& {Roche}, N. 1995, \mnras, 273, 257

\bibitem[{{Momany} {et~al.}(2001){Momany}, {Vandame}, {Zaggia}, {Mignani}, {da
  Costa}, {Arnouts}, {Groenewegen}, {Hatziminaoglou}, {Madejsky}, {Rit{\'e}},
  {Schirmer}, \& {Slijkhuis}}]{momany2001}
{Momany}, Y., {Vandame}, B., {Zaggia}, S., {et~al.} 2001, \aap, 379, 436

\bibitem[{{Nagamine}(2001)}]{nagamine2001}
{Nagamine}, K. 2001, Ph.D.~Thesis

\bibitem[{{Nagashima} {et~al.}(2002){Nagashima}, {Yoshii}, {Totani}, \&
  {Gouda}}]{nagashima2002}
{Nagashima}, M., {Yoshii}, Y., {Totani}, T., \& {Gouda}, N. 2002, \apj, 578,
  675

\bibitem[{{O'Connell}(1999)}]{oconnell1999}
{O'Connell}, R.~W. 1999, \araa, 37, 603

\bibitem[{{Oliver} {et~al.}(2000){Oliver}, {Rowan-Robinson}, {Alexander},
  {Almaini}, {Balcells}, {Baker}, {Barcons}, {Barden}, {Bellas-Velidis},
  {Cabrera-Guerra}, {Carballo}, {Cesarsky}, {Ciliegi}, {Clements}, {Crockett},
  {Danese}, {Dapergolas}, {Drolias}, {Eaton}, {Efstathiou}, {Egami}, {Elbaz},
  {Fadda}, {Fox}, {Franceschini}, {Genzel}, {Goldschmidt}, {Graham},
  {Gonzalez-Serrano}, {Gonzalez-Solares}, {Granato}, {Gruppioni},
  {Herbstmeier}, {H{\' e}raudeau}, {Joshi}, {Kontizas}, {Kontizas},
  {Kotilainen}, {Kunze}, {La Franca}, {Lari}, {Lawrence}, {Lemke},
  {Linden-V{\o}rnle}, {Mann}, {M{\' a}rquez}, {Masegosa}, {Mattila}, {McMahon},
  {Miley}, {Missoulis}, {Mobasher}, {Morel}, {N{\o}rgaard-Nielsen}, {Omont},
  {Papadopoulos}, {Perez-Fournon}, {Puget}, {Rigopoulou}, {Rocca-Volmerange},
  {Serjeant}, {Silva}, {Sumner}, {Surace}, {Vaisanen}, {van der Werf}, {Verma},
  {Vigroux}, {Villar-Martin}, \& {Willott}}]{oliver2000}
{Oliver}, S., {Rowan-Robinson}, M., {Alexander}, D.~M., {et~al.} 2000, \mnras,
  316, 749

\bibitem[{{Peebles}(1982)}]{peebles1982}
{Peebles}, P.~J.~E. 1982, \apjl, 263, L1

\bibitem[{{Persic} \& {Rephaeli}(2002)}]{persic2002}
{Persic}, M. \& {Rephaeli}, Y. 2002, \aap, 382, 843

\bibitem[{{Picard}(1991)}]{picard1991}
{Picard}, A. 1991, \aj, 102, 445

\bibitem[{{Pickles}(1998)}]{pickles1998}
{Pickles}, A.~J. 1998, \pasp, 110, 863

\bibitem[{{Puget} {et~al.}(1996){Puget}, {Abergel}, {Bernard}, {Boulanger},
  {Burton}, {Desert}, \& {Hartmann}}]{puget1996}
{Puget}, J.-L., {Abergel}, A., {Bernard}, J.-P., {et~al.} 1996, \aap, 308, L5

\bibitem[{{Puget} {et~al.}(1999){Puget}, {Lagache}, {Clements}, {Reach},
  {Aussel}, {Bouchet}, {Cesarsky}, {D{\'e}sert}, {Dole}, {Elbaz},
  {Franceschini}, {Guiderdoni}, \& {Moorwood}}]{puget1999}
{Puget}, J.~L., {Lagache}, G., {Clements}, D.~L., {et~al.} 1999, \aap, 345, 29

\bibitem[{{Rieke} {et~al.}(2004){Rieke}, {Young}, {Engelbracht}, {Kelly},
  {Low}, {Haller}, {Beeman}, {Gordon}, {Stansberry}, {Misselt}, {Cadien},
  {Morrison}, {Rivlis}, {Latter}, {Noriega-Crespo}, {Padgett}, {Stapelfeldt},
  {Hines}, {Egami}, {Muzerolle}, {Alonso-Herrero}, {Blaylock}, {Dole}, {Hinz},
  {Le Floc'h}, {Papovich}, {P{\' e}rez-Gonz{\' a}lez}, {Smith}, {Su},
  {Bennett}, {Frayer}, {Henderson}, {Lu}, {Masci}, {Pesenson}, {Rebull}, {Rho},
  {Keene}, {Stolovy}, {Wachter}, {Wheaton}, {Werner}, \&
  {Richards}}]{rieke2004}
{Rieke}, G.~H., {Young}, E.~T., {Engelbracht}, C.~W., {et~al.} 2004, \apjs,
  154, 25

\bibitem[{{Roche} {et~al.}(1996){Roche}, {Shanks}, {Metcalfe}, \&
  {Fong}}]{roche1996}
{Roche}, N., {Shanks}, T., {Metcalfe}, N., \& {Fong}, R. 1996, \mnras, 280, 397

\bibitem[{{Rowan-Robinson} {et~al.}(2004){Rowan-Robinson}, {Lari},
  {Perez-Fournon}, {Gonzalez-Solares}, {La Franca}, {Vaccari}, {Oliver},
  {Gruppioni}, {Ciliegi}, {H{\'e}raudeau}, {Serjeant}, {Efstathiou},
  {Babbedge}, {Matute}, {Pozzi}, {Franceschini}, {Vaisanen}, {Afonso-Luis},
  {Alexander}, {Almaini}, {Baker}, {Basilakos}, {Barden}, {del Burgo},
  {Bellas-Velidis}, {Cabrera-Guerra}, {Carballo}, {Cesarsky}, {Clements},
  {Crockett}, {Danese}, {Dapergolas}, {Drolias}, {Eaton}, {Egami}, {Elbaz},
  {Fadda}, {Fox}, {Genzel}, {Goldschmidt}, {Gonzalez-Serrano}, {Graham},
  {Granato}, {Hatziminaoglou}, {Herbstmeier}, {Joshi}, {Kontizas}, {Kontizas},
  {Kotilainen}, {Kunze}, {Lawrence}, {Lemke}, {Linden-V{\o}rnle}, {Mann},
  {M{\'a}rquez}, {Masegosa}, {McMahon}, {Miley}, {Missoulis}, {Mobasher},
  {Morel}, {N{\o}rgaard-Nielsen}, {Omont}, {Papadopoulos}, {Puget},
  {Rigopoulou}, {Rocca-Volmerange}, {Sedgwick}, {Silva}, {Sumner}, {Surace},
  {Vila-Vilaro}, {van der Werf}, {Verma}, {Vigroux}, {Villar-Martin},
  {Willott}, {Carrami{\~n}ana}, \& {Mujica}}]{rowanrobinson2004}
{Rowan-Robinson}, M., {Lari}, C., {Perez-Fournon}, I., {et~al.} 2004, \mnras,
  351, 1290

\bibitem[{{Rowan-Robinson} {et~al.}(1999){Rowan-Robinson}, {Oliver},
  {Efstathiou}, {Gruppioni}, {Serjeant}, {Cesarsky}, {Danese}, {Franceschini},
  {Genzel}, {Lawrence}, {Lemke}, {McMahon}, {Miley}, {Perez-Fournon}, {Puget},
  {Rocca-Volmerange}, {Ciliegi}, {Heraudeau}, {Surace}, {Lafranca}, \& {ElAIS
  Consortium}}]{rowanrobinson1999}
{Rowan-Robinson}, M., {Oliver}, S., {Efstathiou}, A., {et~al.} 1999, in ESA
  SP-427: The Universe as Seen by ISO, 1011

\bibitem[{{Schlegel} {et~al.}(1998){Schlegel}, {Finkbeiner}, \&
  {Davis}}]{schlegel1998}
{Schlegel}, D.~J., {Finkbeiner}, D.~P., \& {Davis}, M. 1998, \apj, 500, 525

\bibitem[{{Shields}(1978)}]{shields1978}
{Shields}, G.~A. 1978, \nat, 272, 706

\bibitem[{{Silva} {et~al.}(1998){Silva}, {Granato}, {Bressan}, \&
  {Danese}}]{silva1998}
{Silva}, L., {Granato}, G.~L., {Bressan}, A., \& {Danese}, L. 1998, \apj, 509,
  103

\bibitem[{{Space Telescope Science Institute} \& {Osservatorio Astronomico di
  Torino}(2001)}]{gsc2}
{Space Telescope Science Institute}, . \& {Osservatorio Astronomico di Torino}.
  2001, VizieR Online Data Catalog, 1271, 0

\bibitem[{{Spergel} {et~al.}(2003){Spergel}, {Verde}, {Peiris}, {Komatsu},
  {Nolta}, {Bennett}, {Halpern}, {Hinshaw}, {Jarosik}, {Kogut}, {Limon},
  {Meyer}, {Page}, {Tucker}, {Weiland}, {Wollack}, \& {Wright}}]{spergel2003}
{Spergel}, D.~N., {Verde}, L., {Peiris}, H.~V., {et~al.} 2003, \apjs, 148, 175

\bibitem[{{Stone} {et~al.}(1999){Stone}, {Pier}, \& {Monet}}]{stone1999}
{Stone}, R.~C., {Pier}, J.~R., \& {Monet}, D.~G. 1999, \aj, 118, 2488

\bibitem[{{Surace} {et~al.}(2004){Surace}, {Shupe}, {Fang}, {Lonsdale},
  {Gonzalez-Solares}, {Baddedge}, {Frayer}, {Evans}, {Jarrett}, {Padgett},
  {Castro}, {Masci}, {Domingue}, {Fox}, {Rowan-Robinson}, {Perez-Fournon},
  {Olivier}, {Poletta}, {Berta}, {Rodighiero}, {Vaccari}, {Stacey},
  {Hatziminaoglou}, {Farrah}, {Siana}, {Smith}, {Franceschini}, {Owen},
  {Pierre}, {Xu}, {Afonso-Luis}, {Davoodi}, {Dole}, {Pozzi}, {Salaman}, \&
  {Waddington}}]{surace2004}
{Surace}, J.~A., {Shupe}, D.~L., {Fang}, F., {et~al.} 2004, VizieR Online Data
  Catalog, 2255, 0

\bibitem[{{Treu}(2004)}]{treu2004}
{Treu}, T. 2004, in Clusters of Galaxies: Probes of Cosmological Structure and
  Galaxy Evolution, 178

\bibitem[{{Vaccari} {et~al.}(2005){Vaccari}, {Lari}, {Angeretti}, {Fadda},
  {Gruppioni}, {Pozzi}, {Prouton}, {Aussel}, {Babbedge}, {Ciliegi},
  {Franceschini}, {Gonzalez-Solares}, {La Franca}, {Oliver}, {Perez-Fournon},
  {Rowan-Robinson}, {Serjeant}, \& {V{\" a}is{\" a}nen}}]{vaccari2005}
{Vaccari}, M., {Lari}, C., {Angeretti}, L., {et~al.} 2005, \mnras, 358, 397

\bibitem[{{Valdes}(1998)}]{valdes1998}
{Valdes}, F.~G. 1998, in ASP Conf. Ser. 145: Astronomical Data Analysis
  Software and Systems VII, 53

\bibitem[{{Williams} {et~al.}(1996){Williams}, {Blacker}, {Dickinson}, {Dixon},
  {Ferguson}, {Fruchter}, {Giavalisco}, {Gilliland}, {Heyer}, {Katsanis},
  {Levay}, {Lucas}, {McElroy}, {Petro}, {Postman}, {Adorf}, \&
  {Hook}}]{williams1996}
{Williams}, R.~E., {Blacker}, B., {Dickinson}, M., {et~al.} 1996, \aj, 112,
  1335

\bibitem[{{Wilson}(2003)}]{wilson2003}
{Wilson}, G. 2003, \apj, 585, 191

\bibitem[{{Yasuda} {et~al.}(2001){Yasuda}, {Fukugita}, {Narayanan}, {Lupton},
  {Strateva}, {Strauss}, {Ivezi{\' c}}, {Kim}, {Hogg}, {Weinberg}, {Shimasaku},
  {Loveday}, {Annis}, {Bahcall}, {Blanton}, {Brinkmann}, {Brunner}, {Connolly},
  {Csabai}, {Doi}, {Hamabe}, {Ichikawa}, {Ichikawa}, {Johnston}, {Knapp},
  {Kunszt}, {Lamb}, {McKay}, {Munn}, {Nichol}, {Okamura}, {Schneider},
  {Szokoly}, {Vogeley}, {Watanabe}, \& {York}}]{yasuda2001}
{Yasuda}, N., {Fukugita}, M., {Narayanan}, V.~K., {et~al.} 2001, \aj, 122, 1104

\end{thebibliography}

\end{document}